\documentclass{article}
\usepackage[utf8]{inputenc}

\usepackage[english]{babel}	% English language/hyphenation
\usepackage{amsmath,amsfonts,amsthm} % Math packages
\usepackage{graphicx} %[final]
\usepackage[hyphens]{url}

\usepackage[authoryear,round]{natbib}
\usepackage{tabularx}
\newcolumntype{Y}{>{\centering\arraybackslash}X}
\newcolumntype{L}{>{\raggedright\arraybackslash}X}
\newcolumntype{R}{>{\raggedleft\arraybackslash}X}
\usepackage{adjustbox}
\graphicspath{{img/}}
\usepackage[left=4cm,right=4cm,top=3.5cm,bottom=3.5cm]{geometry}
\usepackage{listings}
\usepackage{verbatim}
\usepackage{bigstrut}

\usepackage{fancyvrb}
\usepackage{enumerate}

\usepackage{multirow}
\usepackage{booktabs}
\usepackage{fancyhdr}
\usepackage{float}

\usepackage[left]{eurosym}

\begin{document}

\begin{titlepage}
\setcounter{page}{0}
\begin{flushright}
\end{flushright}

\vskip 2.0cm

\begin{center}

{\LARGE\bf
Cyber bonds and their pricing models}

\vspace{18mm}

{\Large
Oleg Kolesnikov$\,{}^{\dagger}$, \
Alexander Markov$\,{}^{\dagger,\ast}$, \
Daulet Smagulov$\,{}^{\dagger}$
Sergejs Solovjovs$\,{}^{\dagger}$}

\vspace{8mm}

\noindent ${}^\dagger${\em Econophysica Ltd., Annecy Court, Ferry Works, Summer Road, Thames Ditton, Surrey KT7 0QJ, United Kingdom}

\vspace{4mm}

\noindent ${}^\ast${\em Corresponding author}

\vspace{8mm}

\today

\vspace{18mm}

\begin{abstract}
Motivated by the developments in cyber risk treatment in the finance industry, we propose a general framework of cyber bond, whose main purpose is to insure (compensate) losses of a cyber attack. Based on a database of publicly available cyber events, we determine cyber loss distribution parameters and use them to numerically simulate cyber bond price, yield, and other characteristics. We also consider two possible approaches to cyber bond coupon calculation.
\end{abstract}

\end{center}

%\vspace{12mm}
\vfill
\noindent{\sl E-mails:} oleg.kolesnikov@econophysica.com, alexander.markov@econophysica.com,\\ daulet.smagulov@econophysica.com, sergejs.solovjovs@econophysica.com
%\noindent {\sl Number of words:} 8959\\
%\noindent {\sl Number of figures:} 31\\
%\noindent {\sl Number of tables:} 16\\
\end{titlepage}

%%% ----------------------------------------------------------------------

\section{Introduction}

%%% ----------------------------------------------------------------------

\par
The cybersecurity risk is an operational risk to information and technology assets that has consequences affecting the confidentiality, availability, or integrity of information or information systems~\citep{Cebula2010}. The problem of cyber risk is highly relevant today, as hacking techniques advance rapidly (see, e.g.,~\citep{WEC2016}). New ways of bypassing security and new methods of finding vulnerabilities emerge every year. The number of cyber incidents grows by 25\% and the number of companies falling victim increases by 22\% annually~\citep{Hackmgddn}.
The outcomes of such incidents are reputational damage and material losses.

\par
With the evolution of hacking methods, cybersecurity techniques are improving constantly. These techniques are aimed at preventing cyber incidents and safeguarding companies against potential threats. But the problem is that cyberattacks are very abrupt and modeling and predicting this type of risk is complicated. Since the dynamics of cyberattacks are random and varied, full protection against cyber risk cannot be guaranteed due to the impossibility of predicting the methods and goals of an attack and the constant evolution of cyber risk~\citep{Eling2017}. Besides, there is a so-called zero-day problem when some new hacking technique emerges and no methods of defending against it are available. Also, there is no single universal method of assessing organization cyber security. There are certain models (FAIR, IRAM, CyberVaR), Global Cybersecurity Index (GCI), and ISO 27000, but no general standards and ratings have been developed.
Deficiencies in cyber security techniques, lack of observation data, and complexity of loss assessment are also the factors that inhibit the development of cyber risk defense products.

\par
Since managing cyber risk is complicated, a new problem that needs to be solved emerges. Basing on post-defense, the problem is to soften the effect of cyber incidents and try to benefit from them. One of the possible ways to handle the conditions of the problem is to create a model of risk bonds for a corresponding event.

\par
Today, event risk bonds are widely used. One kind of such securities are weather derivatives (see, e.g.,~\citep{Taylor2006,Ahcan2012,Botos2012,Li2018,Stulec2019}). This type of financial instruments can be used by organizations or individuals to prevent risk associated with adverse or unexpected weather conditions as stipulated in the contract. This security is an index-based instrument and uses weather observation data to create an index that a payout can be based on. Another type of such securities are catastrophe bonds, or cat bonds (see, e.g.,~\citep{Longarela1999,Burnecki2005,Braun2015,Nowak2013}). This type of financial instruments helps insurance organizations when a major catastrophe occurs. Such incidents may incur a huge amount of loss since the affected damage cannot be covered by investments~\citep{Aase1999}. Organizations release the above-mentioned cat bonds and pay a coupon to the investor if no catastrophe occurred. Otherwise, the insurance company can stop coupon payments and, in some cases, take out the principal and use the obtained resources to pay their claim holders.

\par
Event risk bonds are usually risky. On the other hand, they have higher coupons. In the case of an event stipulated in the corresponding contract, the trigger is turned on. Depending on the amount of loss and the conditions of the contract, the company can stop coupon payment and take out the contract's notional. This allows the company to offset the potential losses caused by a corresponding event.

\par
In this paper, we explore the issue of cyber post-defense and suggest a new type of financial instruments where cyber incidents are the triggering events. Some organizations and banks already use such securities widely as operational risk bonds that include cyber risk into their scope. However, investors are interested in having a broad selection of securities: one wants more gain, another wants less risk. By restricting the scope of security risk factors to cyber risk only, we allow investors to get a less risky financial instrument, even though less profitable. Considering the rapid evolution of cyber risk, we expect that this type of contract will be highly demanded as a security.

\par
This paper is structured as follows. Section~\ref{pricing:sec1} presents our proposed general methodology of cyber bonds (e.g., their pricing) and provides a short cyber bond example. Section~\ref{distributions:sec1} considers cyber loss distribution fitting. More precisely, given a publicly available cyber event database, Section~\ref{distributions:sec1} provides the respective cyber loss frequency (interval length in days to the next loss) and cyber loss severity (actual loss amount) distributions and their parameters. The former appears to be exponential and the latter -- log-normal. Finally, Section~\ref{example:sec1} provides an extensive example of a cyber bond and estimates its basic characteristics like, e.g., price, yield, \emph{etc.} These characteristics appear to depend significantly on a bond coupon and notional triggers, which (the triggers) determine whether bond coupons and notional are paid to the bondholder. Section~\ref{example:sec1} also considers two approaches to calculating bond coupons, which are based on probability of loss and bond par yield. It also studies the distributions of two specific event groups, namely, cyber-related data breach and cyber-related fraud, and finds out that even though the loss severity for both groups follows the general cyber loss severity distribution, i.e., the log-normal one, the respective loss frequency is different as it follows the non-central Fisher distribution instead of the exponential one.

\section{Methodology of cyber risk bonds\label{pricing:sec1}}
\subsection{Preconditions and concept}

The general concept of event risk securities includes some trigger that stops coupon payments partially or completely.
% This way, an investor can lose both the principal and the coupons.
% Different triggers are used depending on the type of contract and issuers' goals.
% For instance, parametric triggers are sometimes used in cat bonds.
% A parametric trigger considers that coupon payments are stopped when some objective threat level of the coming catastrophe exceeds the initially adjusted threshold even if the catastrophe does not occur.
A trigger takes some factor into account and turns on when the factor surpasses the given threshold.
\cite{Spry2009} describes the following basic trigger types:
\begin{itemize}
    \item with an indemnity trigger, coupon payments are stopped and/or the principal (or its part) is recovered in case one or several base events have their loss amount exceeding the threshold;
    \item with a modeled loss trigger, coupon payments are stopped in case one or several base events have their estimated loss amount exceeding some given threshold; besides, the estimation can be performed constantly while the actual losses are determined sometime after the incident;
    \item with triggers indexed to industry loss, coupon payments are stopped if the total amount of insurance industry losses reaches the given threshold, which is predefined by some competent authority;
    \item a parametric trigger is related to an actual danger and turns on if the critical level of certain conditions is reached, e.\,g. wind speed or area of a forest fire;
    \item a hybrid trigger combines the features of several of the triggers above.
\end{itemize}

\begin{figure}[ht]
    \centering
    \includegraphics[width=\textwidth]{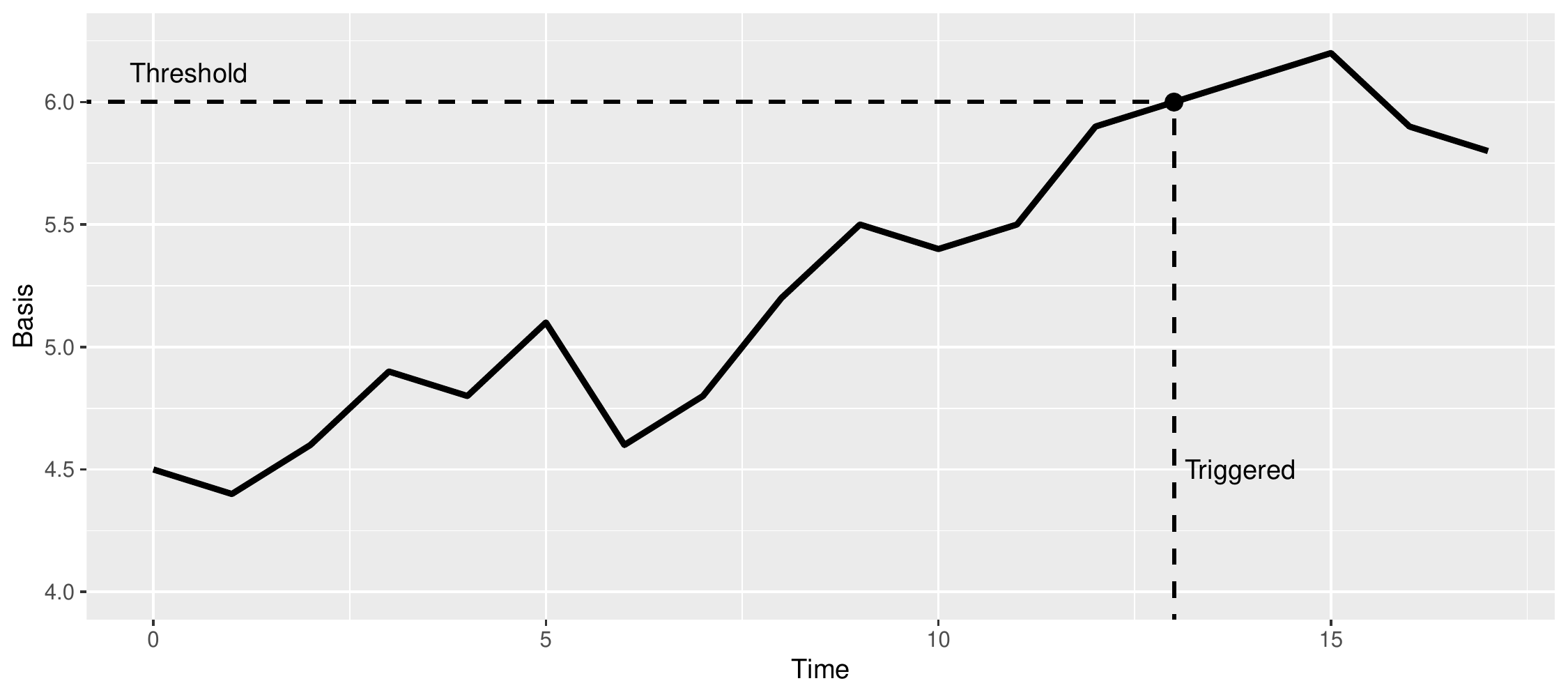}
    \caption{Trigger switches on if the basis exceeds the given threshold}
    \label{fig:trigger}
\end{figure}

A trigger's operating principle is illustrated in Figure~\ref{fig:trigger}. Not all of the trigger types above meet the requirements to be used with cyber risk securities. Today, estimation of loss immediately after a cyberattack is impossible. A certain amount of time is required to determine the damage after the incident. Thus, indemnity and industry loss triggers cannot be used for our securities. Next, due to the unpredictability of cyberattacks, we have no parameters that can signify that a cyber incident is imminent or the probability of its occurrence increases. This makes the parametric trigger inappropriate for the current problem.

On the other hand, possible losses from a cyber incident can be modeled using, e.\,g., Monte-Carlo simulation. Thus, the modeled loss trigger can be chosen for the model. Besides, the calculated threshold of expected losses should be specified in the contract.

Similarly to cat bonds, the structure of financial cyber risk instruments can be represented as follows. An issuer produces securities and takes the principal from an investor. According to the terms of the contract, the issuer then pays coupons to the investor with a given frequency unless the trigger is turned on. Also, the terms of the contract specify whether the principal should be forgiven. If the trigger is not activated for the whole time until the contract maturity, the investor takes all the coupons and the principal. The main problem is the calculation of a cyber risk contract price. The following chapter is dedicated to this problem.

\subsection{Pricing model}
The classical approach to insurance securities pricing is to consider two distributions: events and the periods between them. Within the scope of this article, we are going to follow the same scheme generally, but certain specific features of cyber bonds in terms of pricing should be discussed.

In order to construct the procedure of pricing a contract, we had to
\begin{itemize}
    \item define the set of significant risk factors;
    \item develop a model of the risk factors changing in time;
    \item choose a pricing policy (fair value including risk premium, prudent valuation, etc.) and price valuation approach (analysis, Monte-Carlo, or PDE solution).
\end{itemize}

We considered the standard approach to pricing for event-linked securities: fair value + risk premium.

The risk factor modeling used in the process of pricing is based on the following assumptions. The moments when cyber incidents of type $k$ happen can be described using time intervals between the incidents, where each of the intervals follows the same probability distribution
\begin{equation}
    P(\tau_{k,r}\leq y)=F_k(y),
\end{equation}
where $\tau_{k,r}$ is the interval of time between incidents $r$ and $(r+1)$ of type $k$.
Losses induced by cyber incident $r$ can be described by the probability distribution law
\begin{equation}
    P(\xi_{k,r}\leq x)=G_k(x).
\end{equation}
Distributions $F_k(y)$ and $G_k(x)$ are assumed to be parametric, i.e.
\begin{equation}
    F_k(y)=F_k(y;\theta),\quad G_k(x)=G_k(x;\lambda)
\end{equation}
and model risk is limited by the uncertainty of values of parameter vectors $\theta$ and $\lambda$. This uncertainty is covered by confidence intervals
\begin{equation}
    \theta \in [\theta_l;\theta_u],\quad \lambda \in [\lambda_l;\lambda_u].
\end{equation}
Thus, the price of the contract is defined as the value of the fair price that is the best (maximum) in terms of parameters, where fair price is calculated analytically for linear contracts and using an MC method for the nonlinear ones:
\begin{equation}
    Price=\max_{\theta \in [\theta_l;\theta_u], \lambda \in [\lambda_l;\lambda_u]} FairPrice(F_k(\cdot)=F_k(\cdot;\theta);G_k(\cdot)=G_k(\cdot;\lambda)).
\end{equation}

\subsection{Numerical results}
For the purposes of our study, a Monte Carlo simulation of events and periods between them was carried out. Based on the distributions fitted above, the average price of a bond over a number of simulations was estimated. In addition, Greeks and coupon payment probabilities were calculated.

This cyber risk security pricing model supports the following risk factors:
\begin{itemize}
    \item change of a risk-free rate (risk factor to be taken into account while discounting and the coupon valuation in case it is linked to, for example, LIBOR);
    \item the moment when a cyber incident happened;
    \item aggregated financial losses (optional, can be considered as a single risk factor or a sum of separate risk factors by the source of loss: client payments, business process interruption, reputational loss, etc.);
    \item risk of the ``incorrect" valuation of the risk factors above (model risk, which is significant in case the securities are linked to cyber incidents (insurance or security) due to unavoidable difficulties in modeling).
\end{itemize}

Dependence of the bond value on a risk-free rate and distribution parameters (located on the bounds of confidence intervals) can be assessed using the values of the corresponding Greeks:

\begin{table}[H]
\centering
\begin{tabularx}{\textwidth}
{>{\hsize=1.9\hsize}L >{\hsize=1.25\hsize}L >{\hsize=1.25\hsize}L >{\hsize=.6\hsize}L *{4}{>{\hsize=.75\hsize}R}}
\toprule
Bound & $dS/d\lambda$ &  $dS/d\mu$ &  $dS/d\sigma$ &  $dS/dr$ \\
\midrule
lower & -1.464 & -10.444 & -3.422 &  \quad \bigstrut[b] \\
middle & -2.723 & -11.484 & -6.295 & -2.62 \bigstrut[b] \\
upper & -3.775 & -27.568 & -24.438 & \quad \bigstrut[b] \\

\bottomrule
\end{tabularx}
\caption{Greeks values for 99\% confidence intervals bounds}
\label{tab:greeks}
\end{table}

Presented values are computed using the following values of parameters:
\begin{table}[H]
\centering
\begin{tabularx}{\textwidth}
{>{\hsize=1.9\hsize}L >{\hsize=1.25\hsize}L >{\hsize=1.25\hsize}L >{\hsize=.6\hsize}L *{4}{>{\hsize=.75\hsize}R}}
\toprule
Parameter & Value \\
\midrule
Number of MC iterations & 5000 &  \bigstrut[b]  \\
Trigger for coupons & \$5 billion &  \bigstrut[b]  \\
Trigger for face value & \$50 billion &  \bigstrut[b]  \\
Risk-free rate & 2.65\% &  \bigstrut[b]  \\
Coupon & 30\$, every 182 days  &  \bigstrut[b] \\
Face value & 1000\$ &  \bigstrut[b]  \\
Maturity & 5 years &  \bigstrut[b] \\

\bottomrule
\end{tabularx}
\caption{Other parameters used for Greeks calculation}
\label{tab:params_greeks}
\end{table}

Dependence of the resulting prices on maturity and trigger values is presented in the following figures.

\begin{figure}[H]
    \centering
    \includegraphics[width=0.65\textwidth]{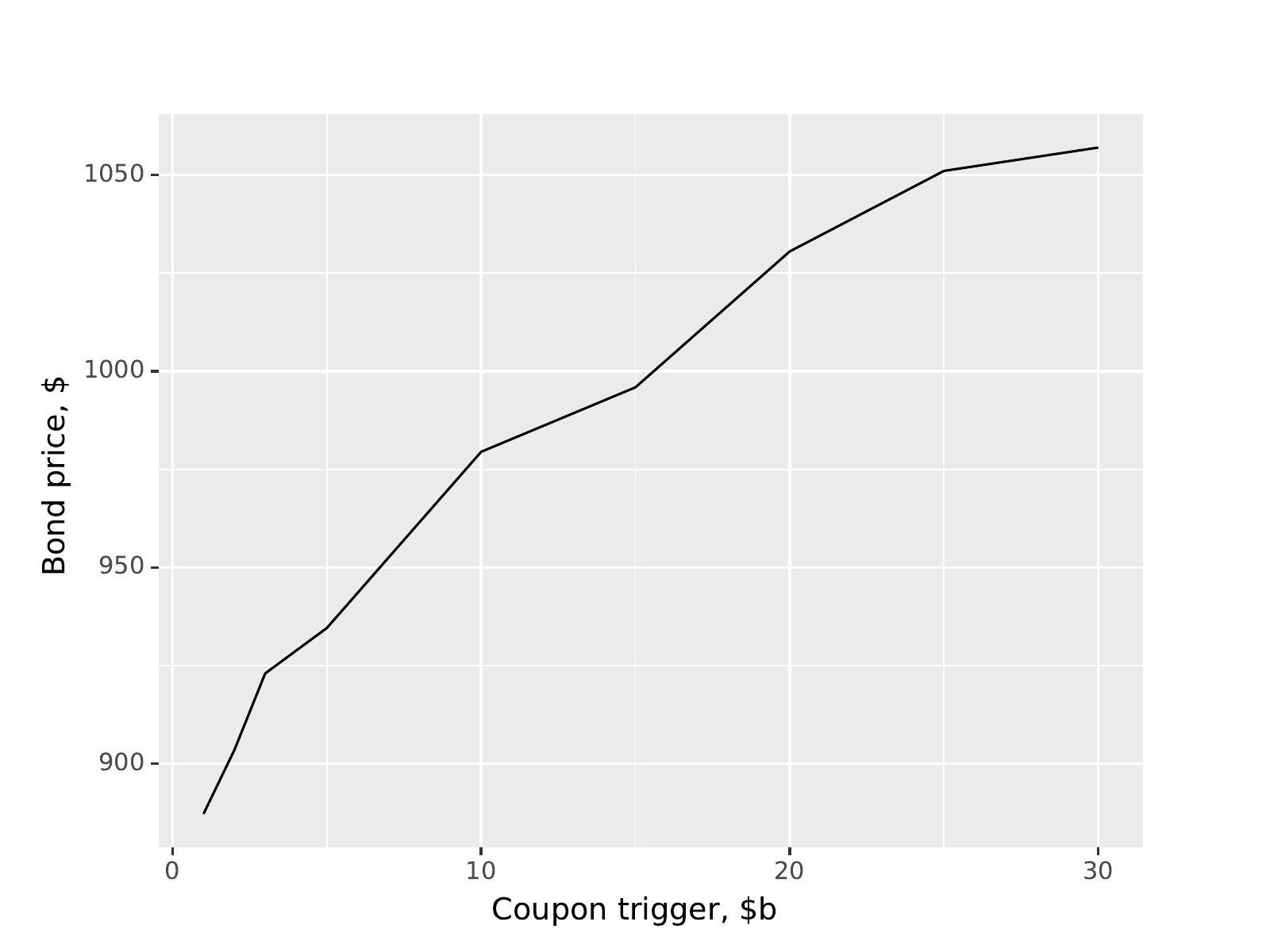}
    \caption{Cyber bond price vs. coupon trigger}
    \label{fig:prices1}
\end{figure}
\begin{figure}[H]
    \centering
    \includegraphics[width=0.65\textwidth]{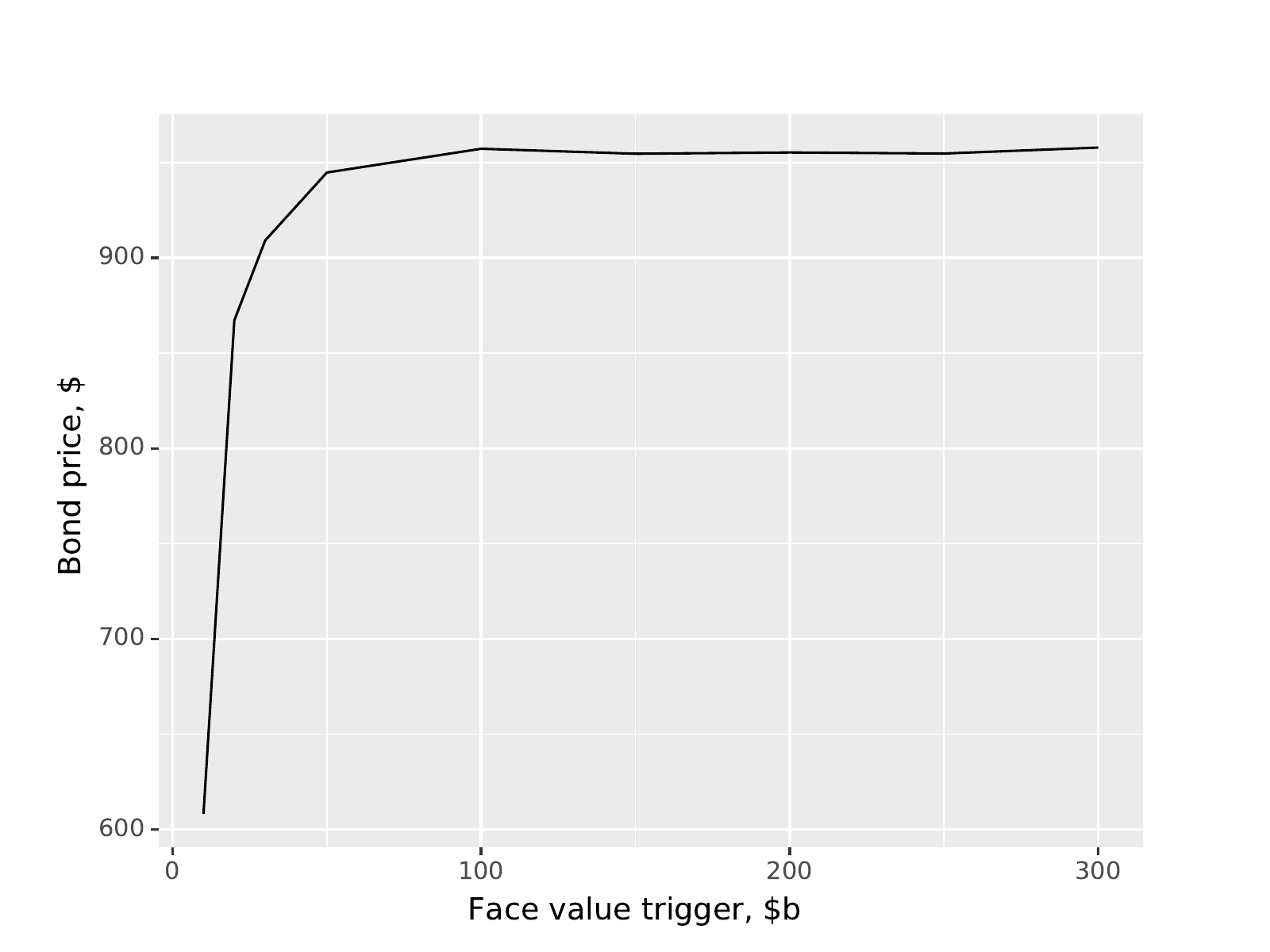}
    \caption{Cyber bond price vs. face value trigger}
    \label{fig:prices2}
\end{figure}

\begin{figure}[H]
    \centering
    \includegraphics[width=0.65\textwidth]{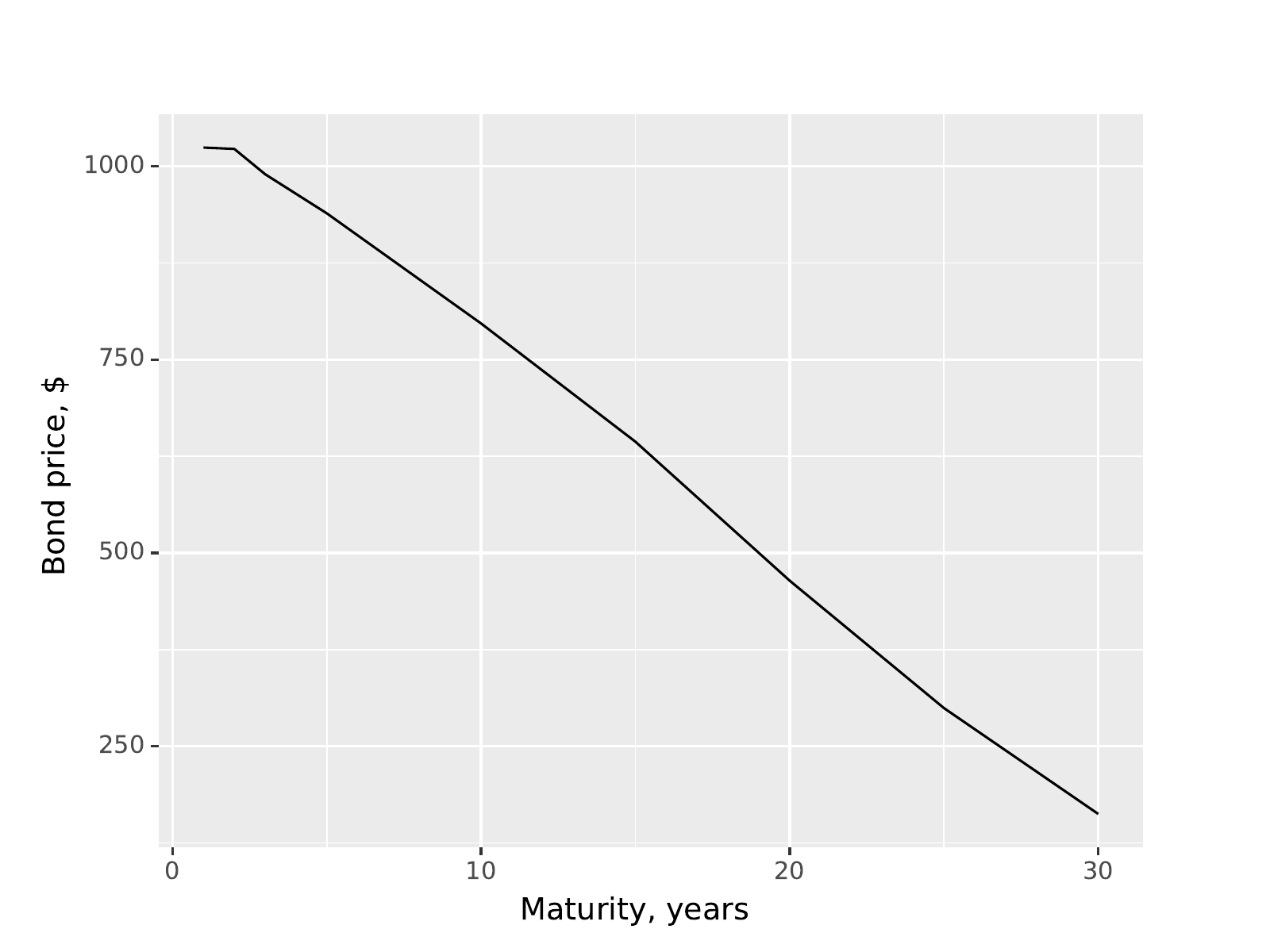}
    \caption{Cyber bond price vs. maturity}
    \label{fig:prices3}
\end{figure}
Figures \ref{fig:prices1}, \ref{fig:prices2} and \ref{fig:prices3} show strong dependence of cyber bond price on triggers and maturity, which is logical, since all the risk can be described by these parameters.

Using parameters from \ref{tab:params_greeks}, the probability of different coupons to be paid could be plotted.
\begin{figure}[H]
    \centering
    \includegraphics[width=0.65\textwidth]{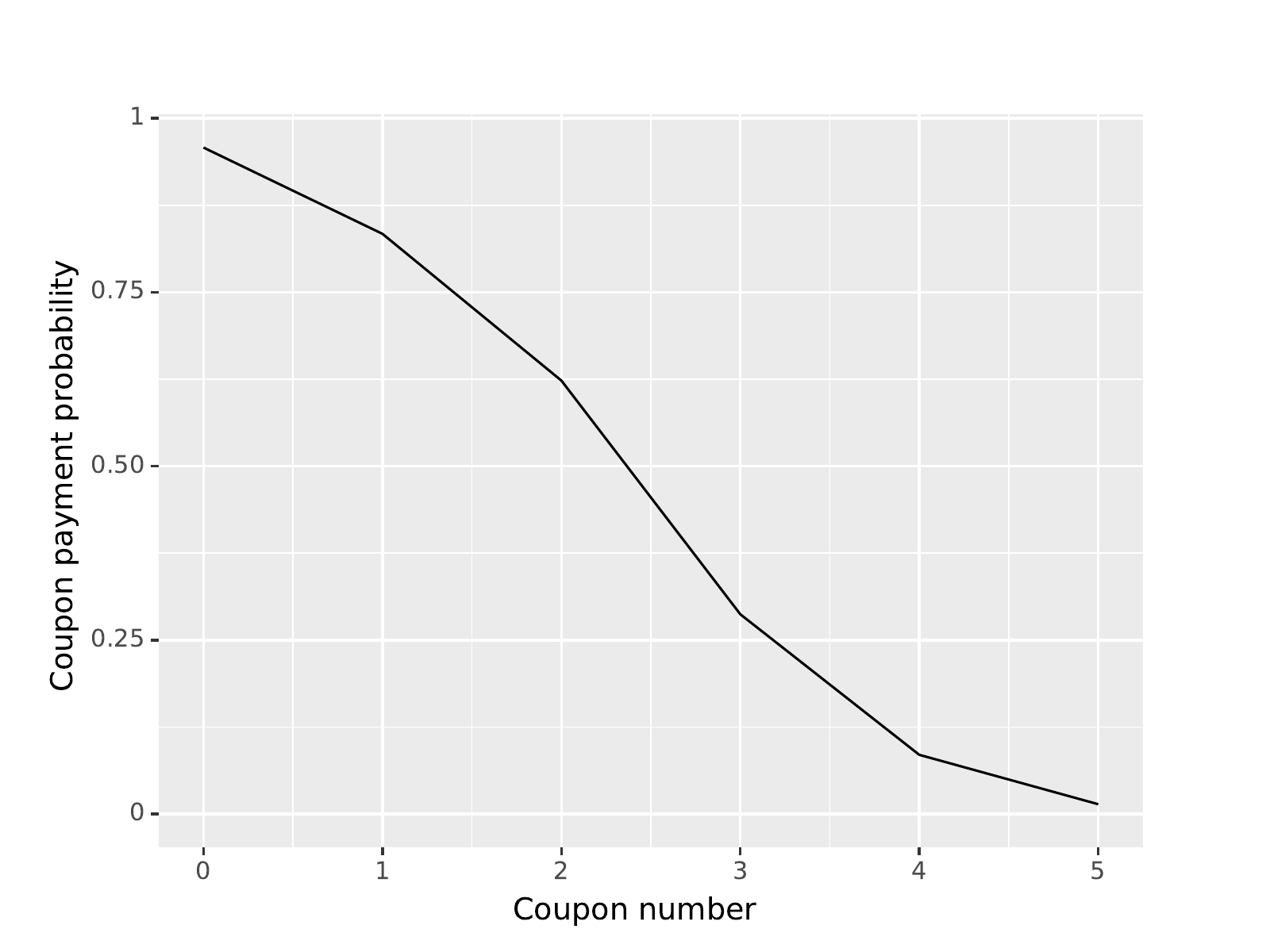}
    \caption{Probability of coupon payment}
    \label{fig:couponProb}
\end{figure}

Payment probabilities shown above display a low survival rate, though it strongly depends on a coupon trigger value.

\section{Example of distribution fitting\label{distributions:sec1}}
\subsection{Historical data}

\begin{figure}[H]
    \centering
    \includegraphics[width=0.49\textwidth]{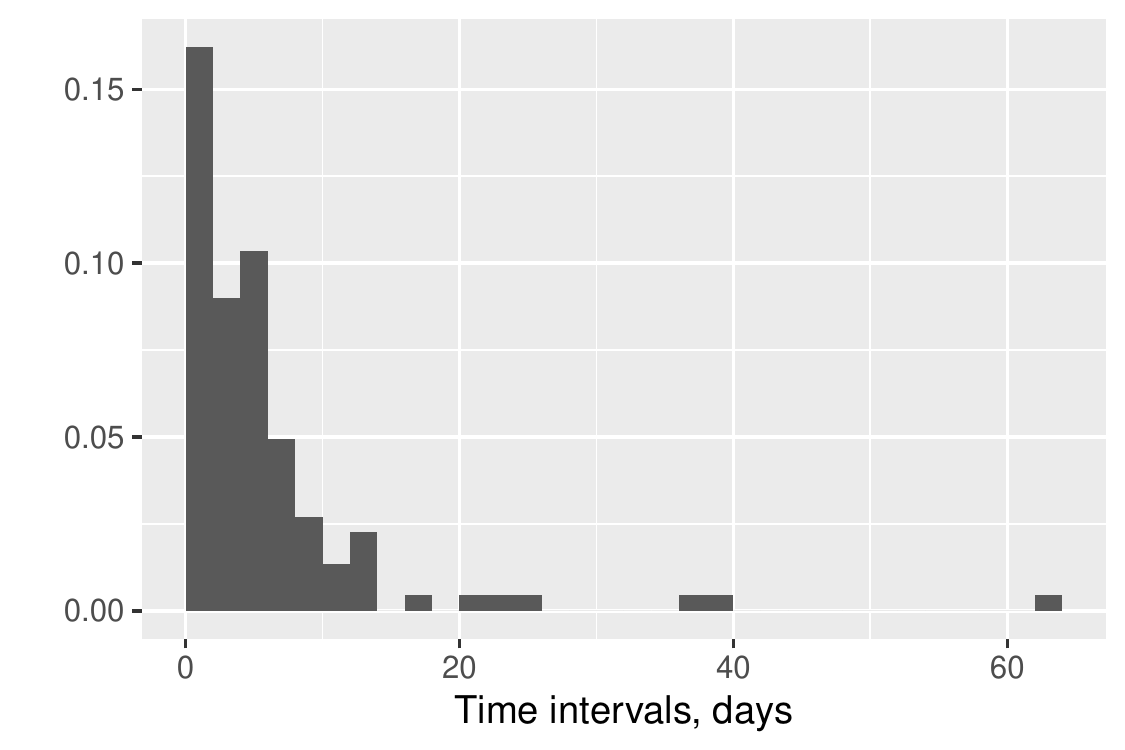}
    \includegraphics[width=0.49\textwidth]{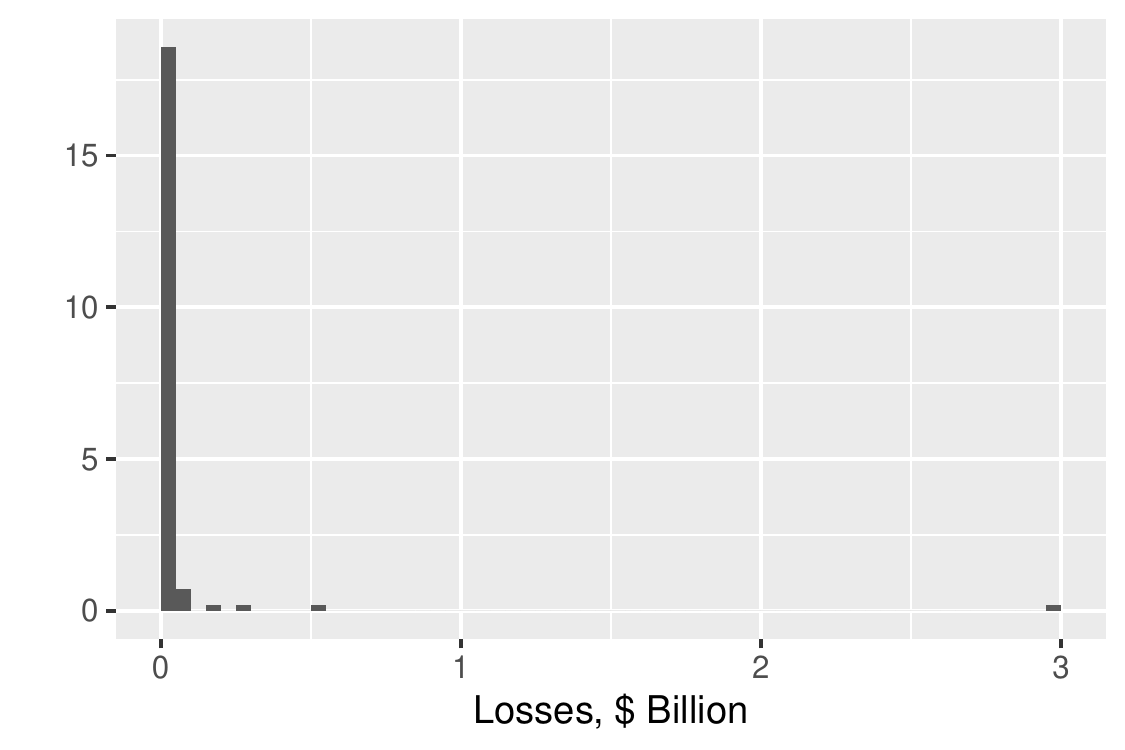}
    \caption{Input data histograms: time intervals between attacks (left) and losses (right)}
    \label{fig:hists}
\end{figure}

According to the previous section, a cyber risk security price depends on the intervals of time between the incidents and the resulting material losses.
In this section, we need to estimate the families of distributions $F_k(y|\theta)$ and $G_k(x|\lambda)$ and their parameters $\theta$ and $\lambda$, respectively.
For this example of estimation, the historical data was downloaded from \citep{Hackmgddn}.
This data contains information about the cyberattacks and cyber incidents that happened within the period of 2017--2018.
Histograms of the data are shown in Figure~\ref{fig:hists}.

\subsection{Estimation and testing}

From the histograms, we conclude that for the data of time series of both intervals and losses, we need to fit the distribution defined on a semi-infinite interval.
It was also considered that the distribution of time intervals should be defined in the zero value.
Thus, Weibull, gamma, $\chi^2$, Fisher, and exponential distributions were selected for fitting.

The maximum likelihood estimation procedure is performed to fit the distribution.
The parameters are estimated using the Nelder\,--\,Mead method.
The results and the corresponding standard error values are represented in Table~\ref{tab:parameters1}.

\begin{table}[H]
\centering
\begin{tabularx}{\textwidth}
{>{\hsize=1.4\hsize}L LLY >{\hsize=.6\hsize}R}
\toprule
Distribution & Parameters & Standard error & Statistic $\omega^2$ & p-value \\
\midrule
Weibull     & $\lambda = 0.334$, $k = 4.07$        & $\sigma_\lambda = 0.03$, $\sigma_k = 1.166$     &  3.94 & < 0.01 \bigstrut \\
Gamma       & $k = 0.233$,       $\theta = 27.403$ & $\sigma_k = 0.024$,      $\sigma_\theta = 6.09$ & 29.58 & < 0.01 \bigstrut \\
$\chi^2$    & $k = 1.02$                           & $\sigma_k = 0.087$                              & 18.04 & < 0.01 \bigstrut \\
Fisher      & $d_1 = 0.392$,     $d_2 = 1.334$     & $\sigma_{d_1} = 0.044$,  $\sigma_{d_2} = 0.298$ & 11.7  & < 0.01 \bigstrut \\
Exponential & $\lambda = 0.156$                    & $\sigma_\lambda = 0.015$                        &  0.23 & 0.22   \bigstrut \\
\bottomrule
\end{tabularx}
\caption{Parameter estimates, standard error, and test results of time interval distribution fitting}
\label{tab:parameters1}
\end{table}

\begin{figure}[H]
    \centering
    \includegraphics[width=\textwidth]{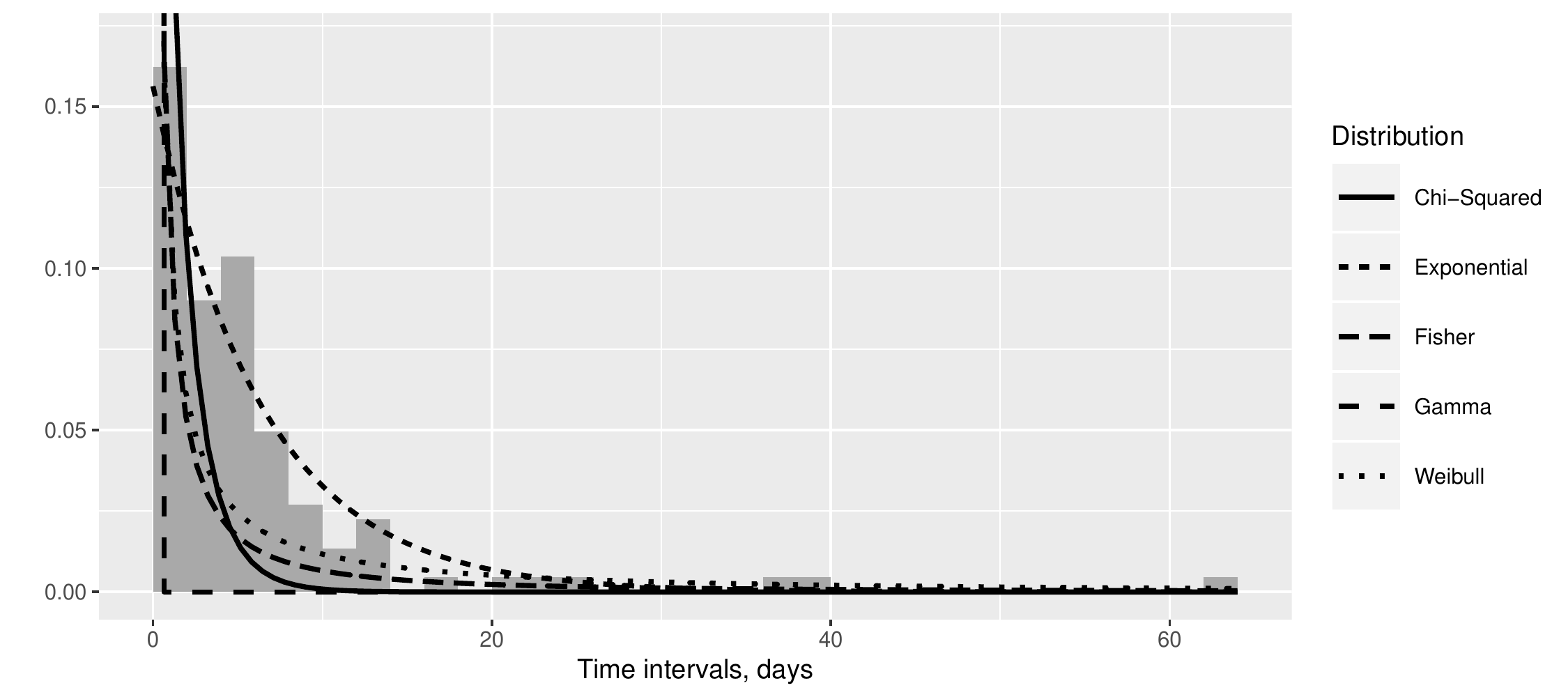}
    \caption{Probability distribution functions selected for time interval fitting}
    \label{fig:fitting1}
\end{figure}

Then, goodness-of-fit testing should be performed for each of the chosen distribution parameters.
For this purpose, the Cram\'er\,--\,von~Mises test is used~\citep{Cramer1928}.
We set the confidence level for the testing to 0.05.
According to the results (Table~\ref{tab:parameters1}), the exponential distribution fits our data better than the others.
All tested probability distribution functions are represented in Figure~\ref{fig:fitting1}.

\begin{figure}[H]
    \centering
    \includegraphics[width=\textwidth]{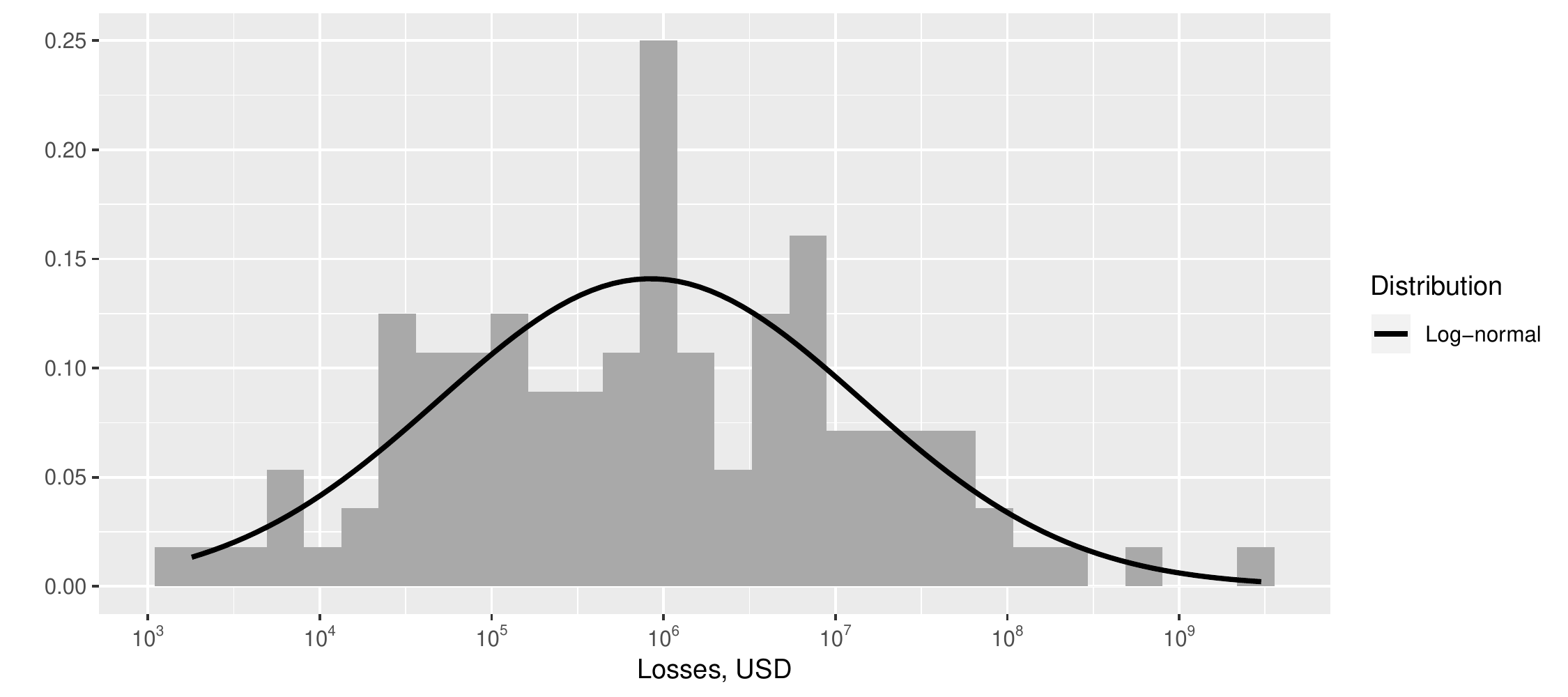}
    \caption{Probability distribution functions selected for loss fitting (logarithmic scale)}
    \label{fig:fitting2}
\end{figure}

Since we consider that cyber incidents always cause some non-zero material losses, we need to take a distribution that is not defined in the zero value.
Also, the corresponding histogram shows that the data has heavy tails (the value close to \$3 billion).
The log-normal distribution meets all the requirements.
\citep{Eling2017} show that the log-normal distribution fits losses data the best.
The estimated parameters and goodness-of-fit test results are shown in Table~\ref{tab:parameters2}.
The test results show that this distribution satisfies the corresponding time series of losses.
The results of fitting are shown in Figure~\ref{fig:fitting2}.
Note that the data is shown in a logarithmic scale, thus the log-normal probability distribution function visually looks like the normal distribution function.

\begin{table}[H]
\centering
\begin{tabularx}{\textwidth}
{>{\hsize=1.4\hsize}L LLY >{\hsize=.6\hsize}R}
\toprule
Distribution & Parameters & Standard error & Statistic $\omega^2$ & p-value \\
\midrule
Log-normal & $\mu = 13.639$, $\sigma = 2.832$ & $\sigma_\mu = 0.268$, $\sigma_\sigma = 0.189$ & 0.038 & 0.94 \\
\bottomrule
\end{tabularx}
\caption{Parameter estimates and standard error for distributions of loss data}
\label{tab:parameters2}
\end{table}

\subsection{Confidence interval for estimated parameters}

Following the idea of prudent valuation methodology~\citep{EBA2014}, we need to calculate the confidence interval for the estimated parameters.
The lower and upper bounds for the parameters were calculated at the 80\%, 97\%, 99\%, and 99.9\% confidence levels.
The values are shown in Table~\ref{tab:conf_interval}.

\begin{table}[H]
\centering
\begin{tabularx}{\textwidth}
{>{\hsize=1.9\hsize}L >{\hsize=1.25\hsize}L >{\hsize=1.25\hsize}L >{\hsize=.6\hsize}L *{4}{>{\hsize=.75\hsize}R}}
\toprule
Distribution & Parameter & Standard error & Bound & 80\% & 97\% & 99\% & 99.9\% \\
\midrule
\multirow{2}{*}{\parbox{\hsize}{\raggedright Exponential (time intervals)}} & \multirow{2}{*}{$\lambda = 0.156$} & \multirow{2}{*}{$\sigma_\lambda = 0.015$} & lower & 0.137 & 0.124 & 0.118 & 0.107 \bigstrut[t] \\
 & & & upper & 0.175 & 0.188 & 0.194 & 0.205 \bigstrut[b] \\
\midrule
\multirow{4}{*}{\parbox{\hsize}{\raggedright Log-normal (losses)}} & \multirow{2}{*}{$\mu = 13.639$} & \multirow{2}{*}{$\sigma_\mu = 0.268$} & lower & 13.296 & 13.058 & 12.950 & 12.758 \bigstrut[t] \\
 & & & upper & 13.982 & 14.220 & 14.328 & 14.519 \bigstrut[b] \\
 & \multirow{2}{*}{$\sigma = 2.832$} & \multirow{2}{*}{$\sigma_\sigma = 0.189$} & lower & 2.589 & 2.421 & 2.344 & 2.209 \bigstrut[t] \\
 & & & upper & 3.074 & 3.243 & 3.319 & 3.455 \bigstrut[b] \\
\bottomrule
\end{tabularx}
\caption{Confidence interval for estimates}
\label{tab:conf_interval}
\end{table}

%%% ----------------------------------------------------------------------

\section{Cyber bond example\label{example:sec1}}

%%% ----------------------------------------------------------------------

%%% ----------------------------------------------------------------------

\subsection{General example setting\label{sec:example1}}

%%% ----------------------------------------------------------------------

\par
In this section, we consider an example of cyber bonds related to a concrete (and randomly chosen) cyber event. As follows from, e.g.,~\citep{MaltaToday2019} in February 2019, a Maltese bank (namely, Bank of Valletta) suffered a cyber attack which saw \EUR{13} million transferred out of the bank through false international transactions. In the wake of the above event, we assume that the bank wants to insure itself against such cyber attacks in the future. It recognizes though that concentrating on information technology security alone (even though it is an essential protection component) could not always be enough since cyber attacks are getting more sophisticated with time. Thus, as a possible additional preventive, one could issue cyber bonds to cover losses in the case of a successful cyber attack against the bank. In the following, we consider an example of such a cyber bond.

\par
Motivated by the above-mentioned cyber attack loss, we will assume that the bond notional is set to \$\,15 million (rounding the result of \EUR/\$ exchange). US dollars are chosen over euros for the specific Federal Reserve (FRED) databases used in the below bond-related calculations. The bond maturity will be chosen as $3$ years (this reflects a time-changing cyber security risk as well as trims large numbers obtained for longer maturities). We assume that the bond pays a coupon every half a year. There could be up to $6$ coupons (we will explain ``could" in a moment) during the life of the bond. From the several possible cyber bond coupon calculation techniques, two of which are considered in Section~\ref{sec:example4}, we choose the par yield approach (described in Section~\ref{sec:example6}) and set bond coupon percentage to $5.09\%$, which amounts to $\$\,764,055.87$. The funding rate for discounting during bond valuation is set to a 3-year Intercontinental Exchange (ICE) swap rate based on USD (taken from~\citep{FRED2019}) which equals $1.52\%$ (as of August $15$, $2019$). The above characteristics of the cyber bond are summarized in Table~\ref{tab:example1}.

\begin{table}[H]
 \centering
 \begin{tabular}[l]{ll}
 \hline
 Bond parameter & Value\\
 \hline\hline
 Notional     & $\$\,15$ million\\
 Maturity     & $3$ years\\
 Coupon rate  & $5.09\%$\\
 Coupon value & $\$\,764,055.87$\\
 Funding rate & $1.52\%$\\
 \hline
 \end{tabular}
 \caption{Cyber bond parameters}\label{tab:example1}
\end{table}

\par
We use the standard formula for bond price $P$:

\begin{equation}
 \label{eq:example1}
 P=\sum\limits_{i=1}^{6} C\cdot e^{-R\cdot\frac{d_i}{365}} + N\cdot e^{-R\cdot\frac{d}{365}},
\end{equation}

\noindent
where $C$ is a coupon value, $R$ is a funding rate, $d_i$ (resp. $d$) is the date of the coupon payment (resp. notional payment) in days, and $N$ is the notional value. In our case, coupon payments are on days 182, 365, 547, 730, 912, and 1095. The last date also corresponds to the payment of the notional.

\par
With Formula~(\ref{eq:example1}) in mind, our cyber bond is assumed to be priced as follows. In the first step, one simulates cyber losses for the bond maturity period, i.e., $3$ years. This simulation includes a day of the loss and its size (cyber loss distribution parameters will be discussed in a moment). In the second step, one uses two triggers, namely, notional and coupon trigger, to determine whether coupons and notional will be paid. More precisely, for every coupon (resp. notional) one sums up the simulated losses up to the respective payment day. If the losses are strictly less than the trigger, then the respective payment is made; otherwise not. This explains our ``could" used to describe the number of coupons of our proposed cyber bond. These triggers are aimed to reflect the state of the industry with respect to cyber attacks. Coupon trigger is assumed to be less than the notional trigger and should define the first level of danger (payment of all or part of the coupons is therefore dropped). Notional trigger represents the second (and highest) level of danger when measures should be taken to compensate losses (payment of the notional is thus dropped). It should be noted, however, that the triggers of other nature are also possible. For example, one could possibly watch for the appearance of a certain number of events with certain losses. One could also watch for the events of a certain nature (e.g. current ransomware attacks) or a certain sequence of cyber events reflecting some pattern. These triggers will not be considered in this paper since they require a deeper and case-specific analysis of an available and comprehensive cyber event database.

%%% ----------------------------------------------------------------------

\subsection{Cyber loss distribution parameters\label{sec:example2}}

%%% ----------------------------------------------------------------------

\par
A word is due to the distribution of cyber losses used in this section. Following the results of Section~\ref{distributions:sec1}, we assume that cyber loses are characterized by two parameters, namely, the interval in days to the next loss and actual loss amount. Interval value is assumed to be distributed exponentially, whereas actual loss value is assumed to have a log-normal distribution. The respective parameters together with the standard errors are given in Table~\ref{tab:example2}.

\begin{table}[H]
 \centering
 \begin{tabular}[l]{lll}
 \hline
 Distribution & Parameters & Standard error\\
 \hline\hline
 Exponential (number of days till the next loss)  & $\lambda = 0.0211$ & $0.0029$\\
 \multirow{2}{*}{Log-normal (actual loss amount)} & $\mu = 14.9179$    & $0.2009$\\
                                                  & $\sigma = 2.3434$  & $0.1421$\\
 \hline
 \end{tabular}
 \caption{Cyber loss distribution parameters}\label{tab:example2}
\end{table}

\noindent
We emphasize that the arrival times of the cyber events are modeled using the exponential distribution. This is similar to the distribution, which underlies models of traded credit instruments such as, e.g., credit default swaps (CDS) (see, e.g.,~\citep{Lando2004}).

\par
The parameters of Table~\ref{tab:example2} were calculated from our obtained table of publicly available cyber events. This table contains 328 items, with 136 of them having publicly disclosed loss amount. These events are ranging from the years 2009~--~2019. The motivating event for our considered cyber bond is taken from this table. The distribution parameters themselves were obtained through the maximum-likelihood method using Nelder--Mead and Broyden--Fletcher--Goldfarb--Shanno optimization algorithms for one- and two-parameter distributions, respectively. Moreover, to estimate the exponential distribution parameter $\lambda$, we made the next two preparatory steps with the available data:

\begin{itemize}
 \item[(S1)] The series of cyber event dates was modified to contain unique items only.
 \item[(S2)] The series of interval lengths between two consecutive cyber events was modified to contain unique items only.
\end{itemize}

\par
Table~\ref{tab:example3} shows the results of the goodness-of-fit (GOF) tests performed by us to justify the choice of cyber loss distribution and the respective distribution parameters (notice that the null hypothesis in each of the tests says that the true distribution is the one mentioned in the first column of Table~\ref{tab:example3} with the respective parameters taken from the second column of Table~\ref{tab:example3}; to reject the null hypothesis with significance level of $5\%$, one searches a value less than $0.05$ in the last column of Table~\ref{tab:example3}).

\begin{table}[H]
 \centering
 \begin{tabular}[l]{llllll}
 \hline
 Distribution & Parameters & GOF test & Test statistic & P-value\\
 \hline\hline
 Exponential                 & $\lambda = 0.0211$ & Chi-square                  & $4.1698$ & $0.8415$ &\\
 Exponential                 & $\lambda = 0.0211$ & Kolmogorov--Smirnov          & $0.1138$ & $0.4642$ &\\
 \hline
 \multirow{2}{*}{Log-normal} & $\mu = 14.9179$    & \multirow{2}{*}{Chi-square} & \multirow{2}{*}{$7.3824$} & \multirow{2}{*}{$0.8313$} &\\
                             & $\sigma = 2.3434$  &                             &          &          &\\
 \multirow{2}{*}{Log-normal} & $\mu = 14.9179$    & \multirow{2}{*}{Kolmogorov--Smirnov} & \multirow{2}{*}{$0.057$} & \multirow{2}{*}{$0.7687$} &\\
                             & $\sigma = 2.3434$  &                             &          &          &\\
 \hline
 \end{tabular}
 \caption{Cyber loss distribution GOF tests}\label{tab:example3}
\end{table}

\begin{figure}[htb]
 \centering
 \includegraphics[width=0.55\textwidth, height=0.55\textwidth]{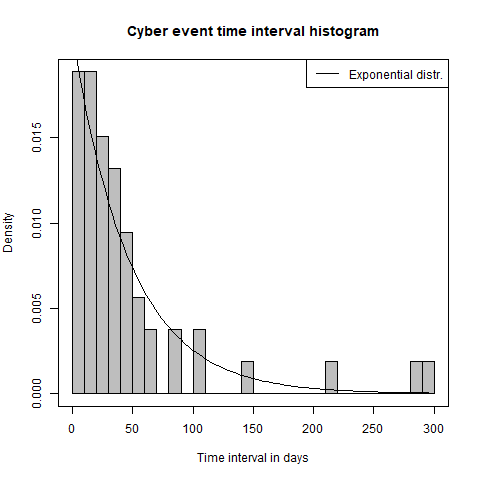}
 \caption{Histogram of time interval (in days) to the next loss and its assumed distribution}
 \label{fig:example0}
\end{figure}

\begin{figure}[htb]
 \centering
 \includegraphics[width=0.55\textwidth, height=0.55\textwidth]{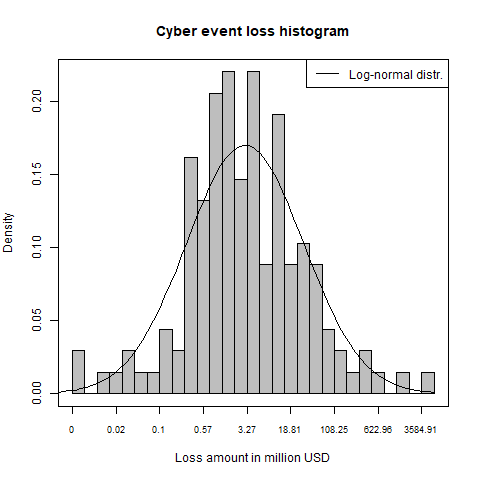}
 \caption{Histogram of actual loss amount and its assumed distribution}
 \label{fig:example0.1}
\end{figure}

\noindent
Additionally, Figures~\ref{fig:example0}, \ref{fig:example0.1} show the histograms of time interval (in days) to the next loss and actual loss amount, both with their assumed distributions.

\par
Backed by the results of Table~\ref{tab:example3} (no rejected null hypothesis) and the visual inspection of Figures~\ref{fig:example0}, \ref{fig:example0.1} (the proposed distributions are ``reasonably" close to the histograms), we will rely on the cyber loss distributions and the parameters of Table~\ref{tab:example2} in the rest of the paper.

\par
Finally, we do not claim to use the all-including table of losses since many cyber events are not publicly disclosed. We do believe, however, that this table is comprehensive enough to provide numerical estimations for our considered example of cyber bonds. Moreover, similar to~\citep{Carter2018}, we do believe that there already exist enough available cyber event data for the successful treatment of cyber bonds.

%%% ----------------------------------------------------------------------

\subsubsection{Distribution of cyber losses of specific type\label{sec:example2.1}}

%%% ----------------------------------------------------------------------

\par
This section answers the question on whether all cyber event types in the used database follow the same distribution. More precisely, we single out two particular cyber event groups (namely, the two biggest ones in our database) and try to find their respective loss frequency (interval length in days to the next loss) and loss severity (actual loss amount) distributions. We will follow the distribution fitting steps of Section~\ref{distributions:sec1}.

%%% ----------------------------------------------------------------------

\paragraph{Cyber-related data breaches:}

%%% ----------------------------------------------------------------------

\par
The first group contains cyber-related data breaches. We notice that this type of cyber events is extremely important for financial institutions often storing sensitive client data, which in their turn could be a lucrative target for cyber criminals. There are altogether $70$ such events in our available database, with $12$ of them having publicly disclosed loss amount.

\par
We first concentrate on loss frequency distribution and its parameters. Table~\ref{tab:example3.1} shows possible loss frequency distributions of cyber-related data breach and their respective parameters together with the standard errors (the reader may recall that we follow the fitting steps of Section~\ref{distributions:sec1} and, thus, the distributions).

\begin{table}[H]
 \centering
 \begin{tabular}[l]{lll}
 \hline
 Distribution & Parameters & Standard error\\
 \hline\hline
 Chi-square               & $k = 24.475$        & $1.1423$\\
 Exponential              & $\lambda = 0.0151$  & $0.0025$\\
 \multirow{3}{*}{Fisher}  & $d_1 = 0.66$        & $0.4092$\\
                          & $d_2 = 2.0643$      & $0.6131$\\
                          & $\lambda = 10.3834$ & $4.3761$\\
 \multirow{2}{*}{Gamma}   & $\alpha = 0.5979$   & $0.1179$\\
                          & $\beta = 0.0091$    & $0.0026$\\
 \multirow{2}{*}{Weibull} & $k = 0.6912$        & $0.0832$\\
                          & $\lambda = 48.5039$ & $12.4018$\\
 \hline
 \end{tabular}
 \caption{Cyber-related data breach loss frequency distribution parameters}\label{tab:example3.1}
\end{table}

\par
The distribution parameters of Table~\ref{tab:example3.1} were obtained through the maximum-likelihood method using Nelder--Mead and Broyden--Fletcher--Goldfarb--Shanno optimization algorithms. The two preparatory steps from the beginning of Section~\ref{sec:example2} ((S1) and (S2)) were taken to adjust the available cyber event occurrence data.

\par
Table~\ref{tab:example3.2} shows our performed GOF tests for the distributions of Table~\ref{tab:example3.1}.

\begin{table}[H]
 \centering
 \begin{tabular}[l]{llllll}
 \hline
 Distribution & Parameters & GOF test & Test statistic & P-value\\
 \hline\hline
 Chi-square               & $k = 24.475$        & Chi square                  & $59.5$ & $0.0000$\\
 Chi-square               & $k = 24.475$        & Kolmogorov--Smirnov          & $0.3479$ & $0.0002$\\
 \hline
 Exponential              & $\lambda = 0.0151$  & Chi square                  & $12.5$ & $0.0853$\\
 Exponential              & $\lambda = 0.0151$  & Kolmogorov--Smirnov          & $0.2743$ & $0.0069$\\
 \hline
 \multirow{3}{*}{Fisher}  & $d_1 = 0.66$        & \multirow{3}{*}{Chi square} & \multirow{3}{*}{$1.5$} & \multirow{3}{*}{$0.9927$}\\
                          & $d_2 = 2.0643$      &                             & &\\
                          & $\lambda = 10.3834$ &                             & &\\
 \multirow{3}{*}{Fisher}  & $d_1 = 0.66$        & \multirow{3}{*}{Kolmogorov--Smirnov} & \multirow{3}{*}{$0.0521$} & \multirow{3}{*}{$0.9999$}\\
                          & $d_2 = 2.0643$      &                             & &\\
                          & $\lambda = 10.3834$ &                             & &\\
 \hline
 \multirow{2}{*}{Gamma}   & $\alpha = 0.5979$        & \multirow{2}{*}{Chi square} & \multirow{2}{*}{$9.5$} & \multirow{2}{*}{$0.1473$}\\
                          & $\beta = 0.0091$   &                             & &\\
 \multirow{2}{*}{Gamma}   & $\alpha = 0.5979$        & \multirow{2}{*}{Kolmogorov--Smirnov} & \multirow{2}{*}{$0.1852$} & \multirow{2}{*}{$0.149$}\\
                          & $\beta = 0.0091$   &                             & &\\
 \hline
 \multirow{2}{*}{Weibull} & $k = 0.6912$        & \multirow{2}{*}{Chi square} & \multirow{2}{*}{$6$} & \multirow{2}{*}{$0.4232$}\\
                          & $\lambda = 48.5039$ &                             & &\\
 \multirow{2}{*}{Weibull} & $k = 0.6912$        & \multirow{2}{*}{Kolmogorov--Smirnov} & \multirow{2}{*}{$0.1385$} & \multirow{2}{*}{$0.4538$}\\
                          & $\lambda = 48.5039$ &                             & &\\
 \hline
 \end{tabular}
 \caption{Cyber-related data breach loss frequency distribution of GOF tests}\label{tab:example3.2}
\end{table}

\par
\begin{figure}[htb]
 \centering
 \includegraphics[width=0.55\textwidth, height=0.55\textwidth]{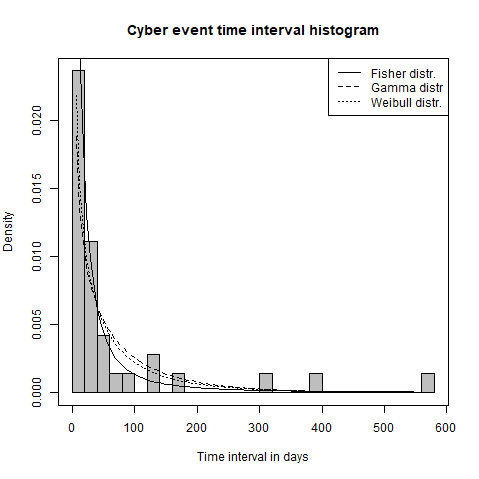}
 \caption{Cyber data breach loss frequency histogram and its assumed distributions}
 \label{fig:example0.3}
\end{figure}

Based on the results of Table~\ref{tab:example3.2}, namely, its last column, where the value below $0.05$ leads to distribution rejection, we assume that the most suitable loss frequency distributions of cyber-related data breach are Fisher, Gamma, and Weibull with the respective parameters. After the visual inspection of Figure~\ref{fig:example0.3}, which displays a cyber-related data breach loss frequency histogram and its assumed distributions, we conclude that the most suitable cyber-related data breach loss frequency distribution is Fisher with $d_1 = 0.66$, $d_2 = 2.0643$, and $\lambda = 10.3834$ (where, $\lambda$ is non-centrality parameter).

\par
We now pay attention to the loss severity distribution and its parameters. Tables~\ref{tab:example3.3}, \ref{tab:example3.4} show cyber-related data breach loss severity distribution parameters with the standard errors and the respective GOF tests.

\begin{table}[H]
 \centering
 \begin{tabular}[l]{lll}
 \hline
 Distribution & Parameters & Standard error\\
 \hline\hline
 \multirow{2}{*}{Log-normal} & $\mu = 15.6826$   & $0.759$\\
                             & $\sigma = 2.6292$ & $0.5367$\\
 \hline
 \end{tabular}
 \caption{Cyber-related data breach loss severity distribution parameters}\label{tab:example3.3}
\end{table}

\begin{table}[H]
 \centering
 \begin{tabular}[l]{llllll}
 \hline
 Distribution & Parameters & GOF test & Test statistic & P-value\\
 \hline\hline
 \multirow{2}{*}{Log-normal} & $\mu = 15.6826$   & \multirow{2}{*}{Chi square} & \multirow{2}{*}{$2$} & \multirow{2}{*}{$0.5724$}\\
                             & $\sigma = 2.6292$ &                             & &\\
 \multirow{2}{*}{Log-normal} & $\mu = 15.6826$   & \multirow{2}{*}{Kolmogorov--Smirnov} & \multirow{2}{*}{$0.1416$} & \multirow{2}{*}{$0.9424$}\\
                             & $\sigma = 2.6292$ &                             & &\\
 \hline
 \end{tabular}
 \caption{Cyber-related data breach loss severity distribution of GOF test}\label{tab:example3.4}
\end{table}

\par
\begin{figure}[htb]
 \centering
 \includegraphics[width=0.55\textwidth, height=0.55\textwidth]{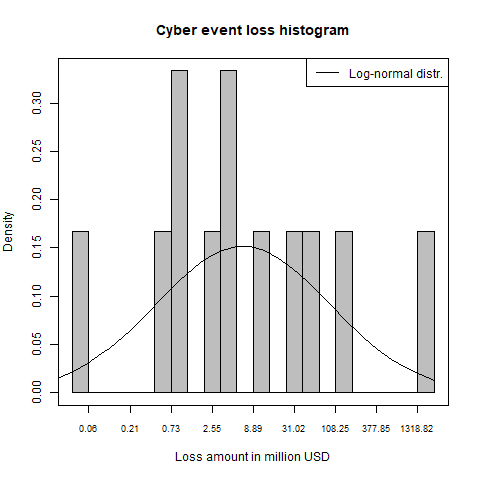}
 \caption{Cyber data breach loss severity histogram and its assumed distribution}
 \label{fig:example0.4}
\end{figure}

Based on the results of Table~\ref{tab:example3.4}, namely, its last column, we assume that the log-normal distribution with the parameters
$\mu = 15.6826$ and $\sigma = 2.6292$ is suitable for cyber-related data breach loss severity. This is confirmed by the visual inspection of Figure~\ref{fig:example0.4}, which displays a cyber-related data breach loss severity histogram and its assumed distribution (notice that the number of events with publicly disclosed loss is small enough and, thus, the respective histogram has a rather awkward shape).

%%% ----------------------------------------------------------------------

\paragraph{Cyber-related fraud:}

%%% ----------------------------------------------------------------------

\par
The second group contains cyber-related fraud events, which are also an important issue for financial institutions (consider, e.g., credit card fraud or e-mail fraud; the latter is gaining in popularity more and more). There are altogether $96$ such events in the available database, with $69$ of them having publicly disclosed loss amount.

\par
We first study loss frequency distribution and its parameters. Table~\ref{tab:example3.5} shows possible cyber-related fraud loss frequency distributions and their respective parameters together with the standard errors.

\begin{table}[H]
 \centering
 \begin{tabular}[l]{lll}
 \hline
 Distribution & Parameters & Standard error\\
 \hline\hline
 Chi-square               & $k = 21.0688$       & $1.0713$\\
 Exponential              & $\lambda = 0.0182$  & $0.0031$\\
 \multirow{3}{*}{Fisher}  & $d_1 = 0.6983$      & $0.4044$\\
                          & $d_2 = 2.5158$      & $0.8009$\\
                          & $\lambda = 10.4261$ & $4.2168$\\
 \multirow{2}{*}{Gamma}   & $\alpha = 0.6109$   & $0.1226$\\
                          & $\beta = 0.0111$    & $0.0033$\\
 \multirow{2}{*}{Weibull} & $k = 0.6983$        & $0.0811$\\
                          & $\lambda = 39.494$  & $10.1437$\\
 \hline
 \end{tabular}
 \caption{Cyber-related fraud loss frequency distribution parameters}\label{tab:example3.5}
\end{table}

\par
The distribution parameters of Table~\ref{tab:example3.5} were obtained through the maximum-likelihood method with the two above-mentioned optimization algorithms. The preparatory steps of Section~\ref{sec:example2} ((S1), (S2)) were taken to adjust the available cyber event occurrence data.

\par
Table~\ref{tab:example3.6} shows GOF tests for the distributions of Table~\ref{tab:example3.5}.

\begin{table}[H]
 \centering
 \begin{tabular}[l]{llllll}
 \hline
 Distribution & Parameters & GOF test & Test statistic & P-value\\
 \hline\hline
 Chi-square               & $k = 21.0688$       & Chi square                  & $56.2857$ & $0.0000$\\
 Chi-square               & $k = 21.0688$       & Kolmogorov--Smirnov          & $0.2839$ & $0.0055$\\
 \hline
 Exponential              & $\lambda = 0.0182$  & Chi square                  & $11.0286$ & $0.1374$\\
 Exponential              & $\lambda = 0.0182$  & Kolmogorov--Smirnov          & $0.2634$ & $0.0124$\\
 \hline
 \multirow{3}{*}{Fisher}  & $d_1 = 0.6983$      & \multirow{3}{*}{Chi square} & \multirow{3}{*}{$1.2571$} & \multirow{3}{*}{$0.996$}\\
                          & $d_2 = 2.5158$      &                             & &\\
                          & $\lambda = 10.4261$ &                             & &\\
 \multirow{3}{*}{Fisher}  & $d_1 = 0.6983$      & \multirow{3}{*}{Kolmogorov--Smirnov} & \multirow{3}{*}{$0.048$} & \multirow{3}{*}{$1$}\\
                          & $d_2 = 2.5158$      &                             & &\\
                          & $\lambda = 10.4261$ &                             & &\\
 \hline
 \multirow{2}{*}{Gamma}   & $\alpha = 0.6109$   & \multirow{2}{*}{Chi square} & \multirow{2}{*}{$8.4571$} & \multirow{2}{*}{$0.2065$}\\
                          & $\beta = 0.0111$    &                             & &\\
 \multirow{2}{*}{Gamma}   & $\alpha = 0.6109$   & \multirow{2}{*}{Kolmogorov--Smirnov} & \multirow{2}{*}{$0.1846$} & \multirow{2}{*}{$0.162$}\\
                          & $\beta = 0.0111$    &                             & &\\
 \hline
 \multirow{2}{*}{Weibull} & $k = 0.6983$        & \multirow{2}{*}{Chi square} & \multirow{2}{*}{$9.4857$} & \multirow{2}{*}{$0.148$}\\
                          & $\lambda = 39.494$  &                             & &\\
 \multirow{2}{*}{Weibull} & $k = 0.6983$        & \multirow{2}{*}{Kolmogorov--Smirnov} & \multirow{2}{*}{$0.1291$} & \multirow{2}{*}{$0.5602$}\\
                          & $\lambda = 39.494$  &                             & &\\
 \hline
 \end{tabular}
 \caption{Cyber-related fraud loss frequency distribution of GOF tests}\label{tab:example3.6}
\end{table}

\par
\begin{figure}[htb]
 \centering
 \includegraphics[width=0.55\textwidth, height=0.55\textwidth]{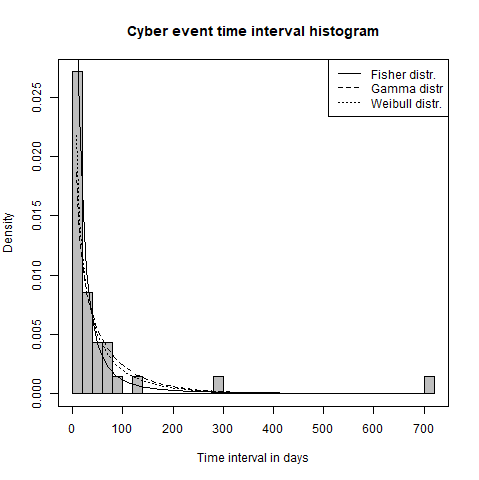}
 \caption{Cyber fraud loss frequency histogram and its assumed distributions}
 \label{fig:example0.5}
\end{figure}

Based on the results of Table~\ref{tab:example3.6}, namely, its last column, where the value below $0.05$ leads to distribution rejection, we assume that the most suitable cyber-related fraud loss frequency distributions are Fisher, Gamma, and Weibull. After the visual inspection of Figure~\ref{fig:example0.5}, a displaying cyber-related data breach loss frequency histogram and its assumed distributions, we conclude that the most suitable cyber-related fraud loss frequency distribution is Fisher with $d_1 = 0.6983$, $d_2 = 2.5158$, and $\lambda = 10.4261$ (non-centrality parameter).

\par
We now concentrate on the loss severity distribution and its parameters. Tables~\ref{tab:example3.7}, \ref{tab:example3.8} show cyber-related fraud loss severity distribution parameters with the standard errors and the respective GOF tests.

\begin{table}[H]
 \centering
 \begin{tabular}[l]{lll}
 \hline
 Distribution & Parameters & Standard error\\
 \hline\hline
 \multirow{2}{*}{Log-normal} & $\mu = 14.6305$   & $0.2959$\\
                             & $\sigma = 2.4582$ & $0.2093$\\
 \hline
 \end{tabular}
 \caption{Cyber-related fraud loss severity distribution parameters}\label{tab:example3.7}
\end{table}

\begin{table}[H]
 \centering
 \begin{tabular}[l]{llllll}
 \hline
 Distribution & Parameters & GOF test & Test statistic & P-value\\
 \hline\hline
 \multirow{2}{*}{Log-normal} & $\mu = 14.6305$   & \multirow{2}{*}{Chi square} & \multirow{2}{*}{$9.5942$} & \multirow{2}{*}{$0.2947$}\\
                             & $\sigma = 2.4582$ &                             & &\\
 \multirow{2}{*}{Log-normal} & $\mu = 14.6305$   & \multirow{2}{*}{Kolmogorov-Smirnov} & \multirow{2}{*}{$0.0993$} & \multirow{2}{*}{$0.504$}\\
                             & $\sigma = 2.4582$ &                             & &\\
 \hline
 \end{tabular}
 \caption{Cyber-related fraud loss severity distribution GOF test}\label{tab:example3.8}
\end{table}

\par
\begin{figure}[htb]
 \centering
 \includegraphics[width=0.55\textwidth, height=0.55\textwidth]{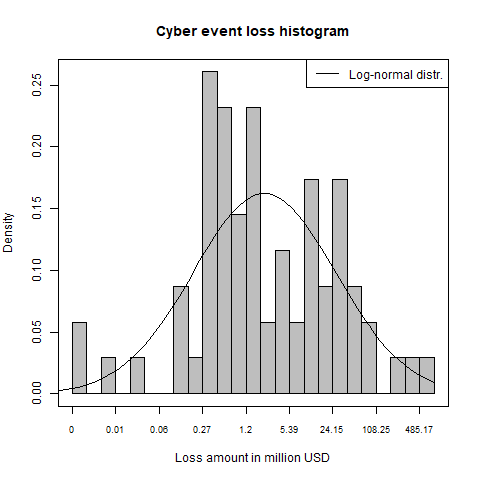}
 \caption{Cyber fraud loss severity histogram and its assumed distribution}
 \label{fig:example0.6}
\end{figure}

Based on the results of Table~\ref{tab:example3.8} (namely, its last column), we assume that the log-normal distribution with the parameters
$\mu = 14.6305$ and $\sigma = 2.4582$ is suitable for cyber-related fraud loss severity. This is confirmed by the visual inspection of Figure~\ref{fig:example0.6}, which displays a cyber-related fraud loss severity histogram and its assumed distribution.

\par
In conclusion, we shall notice that even though loss severity for both the initial database and two checked event groups can be assumed to follow log-normal distribution, loss frequency distribution for the initial (exponential) database differs from that of the two considered event groups (non-central Fisher). Therefore, a particular attention to the loss frequency distribution should be paid, when working with cyber events of a specific type.

%%% ----------------------------------------------------------------------

\subsection{Cyber bond numerical results\label{sec:example3}}

%%% ----------------------------------------------------------------------

\par
This subsection lists the obtained numerical results on our proposed cyber bond example. We rely on Monte Carlo simulations, with the simulation number always being $5,000$.

\par
\begin{figure}[htb]
 \centering
 \includegraphics[width=0.55\textwidth, height=0.55\textwidth]{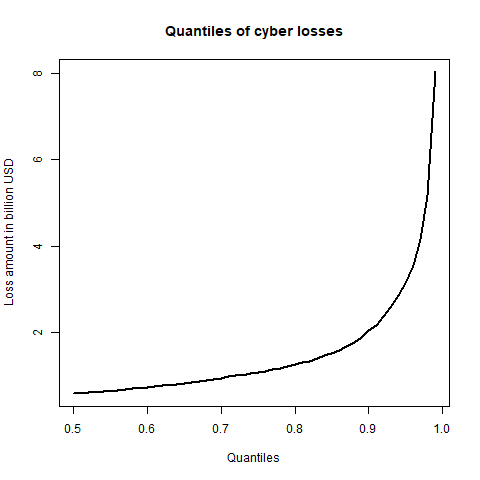}
 \caption{Quantiles of cyber losses for the period of 3 years}
 \label{fig:example1}
\end{figure}

Figure~\ref{fig:example1} represents quantiles of losses for the maturity period of the cyber bond, that is, $3$ years. The quantiles are calculated from Monte Carlo simulations. As can be seen from Figure~\ref{fig:example1}, using our generated loss distribution parameters, the losses can reach the level of $\$\,8$ billion. We use these obtained loss amounts to test the influence of bond coupon and notional triggers on the actual bond parameters (e.g., price, yield,~\emph{etc.}).

\par
\begin{figure}[htb]
 \centering
 \includegraphics[width=0.55\textwidth, height=0.55\textwidth]{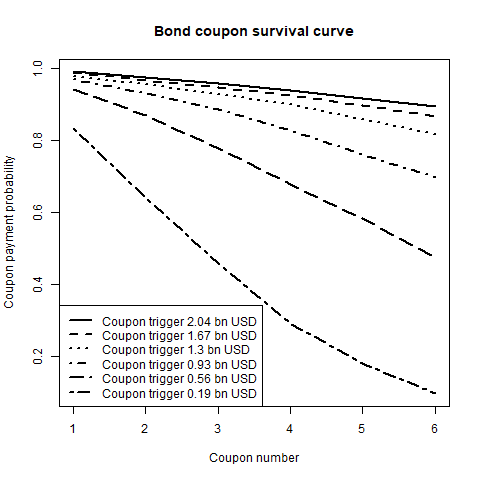}
 \caption{Cyber bond coupon survival curve for 3 years}
 \label{fig:example2}
\end{figure}

We have already mentioned that a bond coupon trigger can influence the number of paid coupons on the bond. We now calculate a bond coupon survival curve. As we mentioned before, the bond is assumed to pay a coupon every $6$ months. Thus, altogether there are $6$ coupons. Figure~\ref{fig:example2} shows the probability of the payment of each coupon depending on the coupon trigger. Recall that a coupon is not paid as soon as the total losses up to the coupon payment date exceed a specific amount (bond coupon trigger). Coupon payment probability is calculated as follows. In the first step, Monte Carlo simulations give a series or $0$s and $1$s depending on whether the respective coupon was paid or not. In the second step, one calculates the probability of coupon payment as the number of $1$s divided by the number of simulations. As can be seen from Figure~\ref{fig:example2}, the coupon trigger of $\$\,2.04$ billion, which corresponds to about $90\%$ quantile as per Figure~\ref{fig:example1}, gives a ``reasonable" coupon payment probability close to $1$. The lines in Figure~\ref{fig:example2} correspond to $10\%$~--~$90\%$ loss quantiles with the uniform step chosen between them.

\par
\begin{figure}[htb]
 \centering
 \includegraphics[width=0.55\textwidth, height=0.55\textwidth]{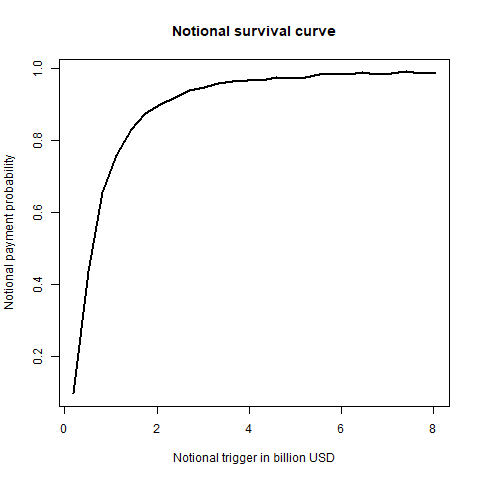}
 \caption{Cyber bond notional survival curve for 3 years}
 \label{fig:example3}
\end{figure}

Similar to the coupon survival curve, we now show a notional survival curve for our proposed cyber bond. As mentioned before, the bond notional is paid back provided that the total loss up to the payment date does not exceed the bond notional trigger. Figure~\ref{fig:example3} shows the probability of a notional payment for different notional triggers. The calculation methodology follows the one for the bond coupons. Notional triggers correspond to loss $10\%$~--~$99\%$ quantiles. Following Figures~\ref{fig:example1}, \ref{fig:example3}, losses above \$\,2 billion (i.e., above $90\%$ quantile) show a ``reasonable" notional payment probability (above $90\%$ as per Figure~\ref{fig:example3}). In general, it is up to a risk-taker to decide which notional payment probability and, therefore, which notional trigger to choose for the issue of the respective cyber bond.

\par
\begin{figure}[htb]
 \centering
 \includegraphics[width=0.55\textwidth, height=0.55\textwidth]{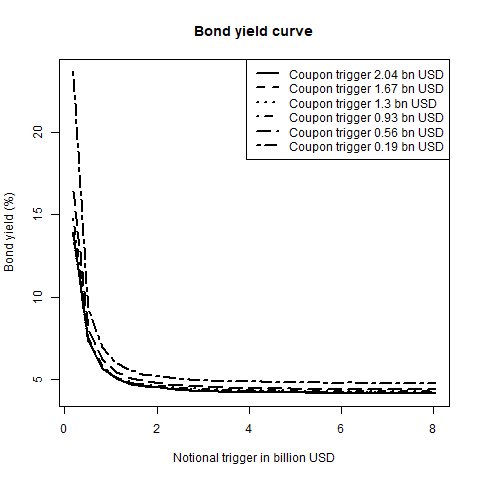}
 \caption{Cyber bond yield curve}
 \label{fig:example4}
\end{figure}

In the next step, we are going to show our proposed cyber bond yield curve, which allows one to judge the profitability of the proposed bond. A bond yield $Y$ is calculated by the following formula:

\begin{equation}
 \label{eq:example2}
 Y = \frac{C}{P}\cdot 100\%,
\end{equation}

\noindent
where $C$ is the bond coupon and $P$ is the bond price. Figure~\ref{fig:example4} shows a cyber bond yield curve depending on both notional as well as coupon triggers. At least two things can be seen from the figure. First, a bond yield is not much influenced by a coupon trigger except for very small notional and coupon triggers. Second, starting from the notional trigger of about $\$\,2$ billion (corresponding roughly to $90\%$ quantile as per Figure~\ref{fig:example1}), the bond yield stabilizes around $5\%$ with the exception of a relatively small coupon trigger of $\$\,0.19$ billion, for which the yield stays strictly above $5\%$, i.e., almost triples our assumed funding rate.

\par
\begin{figure}[htb]
 \centering
 \includegraphics[width=0.55\textwidth, height=0.55\textwidth]{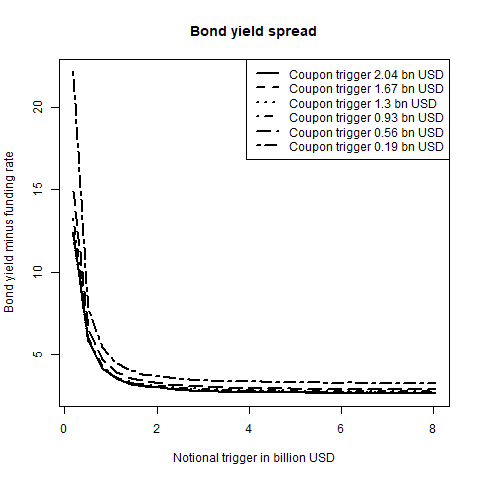}
 \caption{Cyber bond yield spread}
 \label{fig:example5}
\end{figure}

For the reader's convenience, Figure~\ref{fig:example5} shows our proposed cyber bond yield spread, i.e., the difference between the cyber bond yield and the funding rate used in the bond estimate (recall that the assumed funding rate amounts to $1.52\%$) depending on both coupon and notional triggers. As can be seen from the figure, the benefit of investing into our proposed cyber bond is about $3\%$ (under ``reasonable" assumptions on bond triggers).

\par
\begin{figure}[htb]
 \centering
 \includegraphics[width=0.55\textwidth, height=0.55\textwidth]{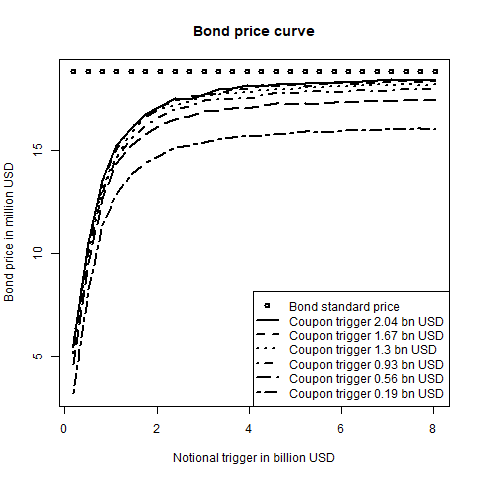}
 \caption{Cyber bond price curve}
 \label{fig:example6}
\end{figure}

Finally, we present our proposed cyber bond price curve. Figure~\ref{fig:example6} shows that for ``reasonable" coupon and notional triggers the bond price stabilizes somewhere around \$\,17 million. We notice that the standard bond price (calculated by Formula~(\ref{eq:example1})), i.e., omitting both a coupon and a notional trigger, is $\$\,18,797,813.26$.

%%% ----------------------------------------------------------------------

\subsection{Cyber bond risk premium\label{sec:example4}}

%%% ----------------------------------------------------------------------

\par
This section considers several approaches to calculating our proposed cyber bond risk premium, namely, a bond coupon, to account for possible coupon (or even the notional itself) loss. It is up to the ultimate issuer of the cyber bond to choose the most suitable technique or invent a new one if necessary.

%%% ----------------------------------------------------------------------

\subsubsection{First alternative: probability of loss\label{sec:example5}}

%%% ----------------------------------------------------------------------

\par
As the first approach to calculating cyber bond coupons, we take the analogy of catastrophe bonds (\emph{cat bonds} for short) studied in, e.g.,~\citep{Ciumas2015}, since a cyber attack could be considered as a kind of catastrophe. A general coupon calculation formula for cat bonds can be written as follows:

\begin{equation}
 \label{eq:example3}
 Coupon\;rate\,(\%) = LIBOR\,(\%) + Risk\;premium\,(\%),
\end{equation}

\noindent
where the risk premium should hedge the exposure of investors (into the bonds) to catastrophe risk (notice that following the current trend (see, e.g.,~\citep{EY2018}), LIBOR rate could be replaced by an alternative reference rate (ARR)). The above-mentioned risk premium could be then determined as follows:

\begin{equation}
 \label{eq:example4}
 Risk\;premium\,(\%) = Constant\,(\%) + Loss\;multiplier \cdot Expected\;loss\,(\%),
\end{equation}

\noindent
where the expected loss is a percentage of the notional expected to be lost during the bond maturity period, the constant is the rate of return requested by investors, and the loss multiplier reflects the uncertainty related to the expected loss. For example, following the results of~\citep[p.~1491]{Ciumas2015}, which are based on catastrophe bonds issued during the years $2006$~-~$2012$ and covering earthquake risks, the constant amounts to $3.35\%$ and the multiplier amounts to $1.4817$. Additionally, as follows from the results of~\citep[p.~168]{Bodoff2013} based on the US catastrophe bonds issued during the years $1998$~-~$2008$ and covering wind risks, the constant (resp. multiplier) amounts to $3.33\%$ (resp. $2.4$).

\par
To simplify the setting, we assume that the constant (the loss multiplier, respectively) in Formula~(\ref{eq:example4}) is equal to $0$ (resp. $1$) since these parameters are highly dependent on investors. Moreover, following~\citep[p.~6]{Braun2014}, we assume that expected loss \emph{EL} is related to probability of loss \emph{PL} and percentage of notional lost \emph{PNL} given the loss occurs as follows:

\begin{equation}
 \label{eq:example5}
EL\,(\%) = PL \cdot PNL\,(\%).
\end{equation}

\par
In our cyber bond setting, \emph{PNL} always amounts to $100\%$, since the notional is either paid back or not paid at all (cases of the partial repayment of the bond notional, despite the fact that they are possible, are not considered in this paper). Thus, in our considered case of cyber bonds, Formula~(\ref{eq:example3}) simplifies to

\begin{equation}
 \label{eq:example6}
Coupon\;rate\,(\%) = LIBOR\,(\%) + PL\,(\%).
\end{equation}

\begin{figure}[htb]
 \centering
 \includegraphics[width=0.55\textwidth, height=0.55\textwidth]{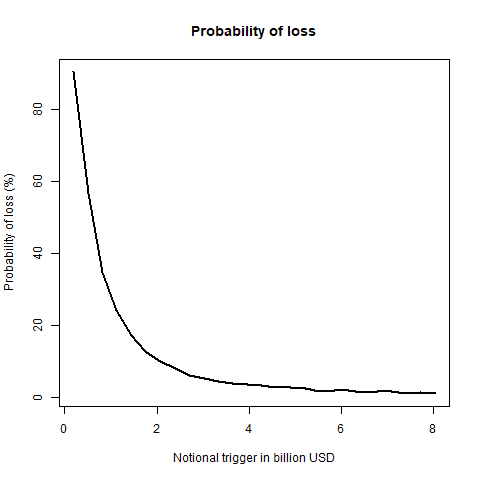}
 \caption{Cyber bond probability of loss for the period of 3 years}
 \label{fig:example7}
\end{figure}

Moreover, it is clear that \emph{PL} of our proposed cyber bond depends on the notional trigger. The lower is the trigger, the more probable it is that the notional will not be paid back. Since this paper does not specify a specific trigger but rather considers trigger influence on bond characteristics, we calculated the average probability of \emph{not} getting back notional from the notional survival curve of Figure~\ref{fig:example3}. Thus, we obtain $11.58\%$ (notice that according to~\citep[p.~1490]{Ciumas2015}, for catastrophe bonds issued during the years $1999$~-~$2012$ and covering earthquake risks, mean, maximum, and minimum risk premiums are $5.63\%$, $14.5\%$, and $1.5\%$, respectively, with the standard deviation being $2.70\%$. Therefore, our obtained number is in line with, e.g., earthquake risk). The average was taken over all the loss quantiles considered in the figure (see Section~\ref{sec:example3} for more details). For convenience of the reader, Figure~\ref{fig:example7} shows \emph{PL} for different notional triggers. As can be seen from the figure, for low notional triggers \emph{PL} could exceed $80\%$.

\par
Altogether, taking into account that $6$-month LIBOR on USD (taken from~\citep{FRED2019} on August $15$, $2019$) is $2.05\%$ and that our proposed cyber bond pays a coupon every $6$ months, the total bond coupon resulting from the first coupon calculation technique (i.e., according to Formula~(\ref{eq:example6})) is $13.63\%$.

%%% ----------------------------------------------------------------------

\subsubsection{Second alternative: par yield\label{sec:example6}}

%%% ----------------------------------------------------------------------

\par
As the second approach to cyber bond coupon calculation, we consider the concept of a bond par yield. Recall from, e.g.,~\citep[p.~85]{Hull2018} that a bond \emph{par yield} is a coupon rate for which the bond price equals its par value, namely, the notional value. It is easy to see (taking into consideration Formula~(\ref{eq:example1})) that the bond coupon $C$ for par yield can be calculated as follows (keeping in mind that $N$ is bond notional):

\begin{equation}
 \label{eq:example7}
 C=\frac{N\cdot(1 - e^{-R\cdot\frac{d}{365}})}{\sum\limits_{i=1}^{6} e^{-R\cdot\frac{d_i}{365}}}.
\end{equation}

\noindent
The respective par yield \emph{PY} is then derived as

\begin{equation}
 \label{eq:example8}
 PY = \frac{C}{N}\cdot 100\% = \frac{(1 - e^{-R\cdot\frac{d}{365}})}{\sum\limits_{i=1}^{6} e^{-R\cdot\frac{d_i}{365}}} \cdot 100\%.
\end{equation}

\noindent
The obtained par yield \emph{PY} is then set to be the bond coupon rate.

\par
In the case of our proposed cyber bond, some of the coupons and even the notional itself may not be paid back to the bondholder, which depends on the actual cyber losses. Thus, we first simulate cyber losses for the bond maturity period and then calculate cyber bond $PY$ by Formula~(\ref{eq:example8}), where now some of $ e^{-R\cdot\frac{d_i}{365}}$ and even $e^{-R\cdot\frac{d}{365}}$ may be taken $0$ depending on whether the respective item is paid to the bondholder or not.

\par
\begin{figure}[htb]
 \centering
 \includegraphics[width=0.55\textwidth, height=0.55\textwidth]{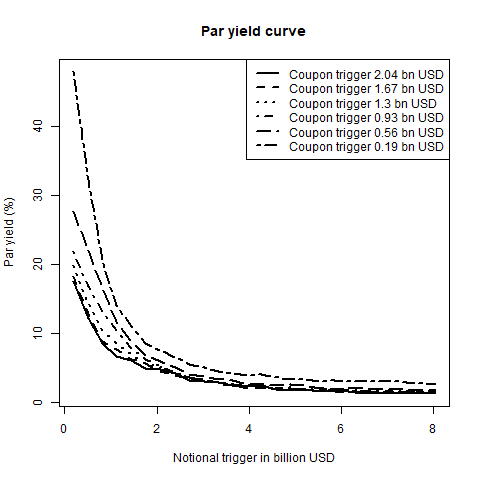}
 \caption{Cyber bond par yield curve}
 \label{fig:example8}
\end{figure}

Figure~\ref{fig:example8} shows the obtained par yield curves depending on both notional and coupon triggers. Since none of the triggers (both depending on the actual bond issuer) is fixed in this paper, we calculate the final coupon percentage as the average par yield over all the obtained curves getting thus $5.09\%$ (recall from Section~\ref{sec:example1} that this amounts to $\$\,764,055.87$ as per assumed bond notional of $\$\,15$ million).

\par
\begin{figure}[htb]
 \centering
 \includegraphics[width=0.55\textwidth, height=0.55\textwidth]{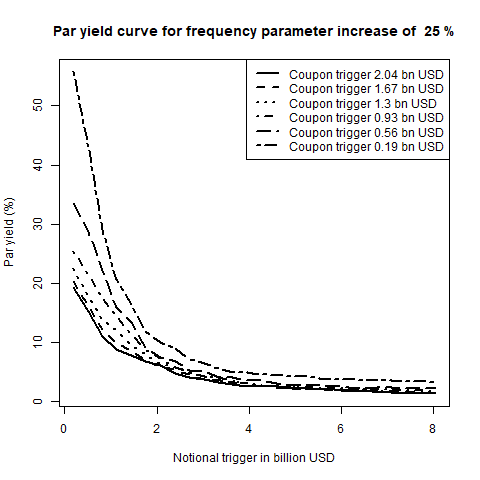}
 \caption{Cyber bond par yield curve for loss frequency parameter $\lambda$ increase of $25\%$}
 \label{fig:example9}
\end{figure}
\begin{figure}[htb]
 \centering
 \includegraphics[width=0.55\textwidth, height=0.55\textwidth]{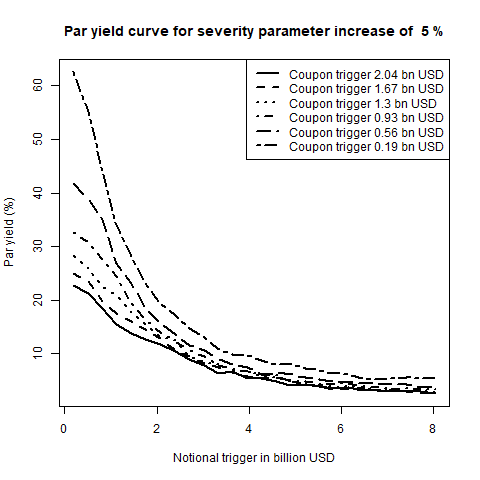}
 \caption{Cyber bond par yield curve for loss severity parameter $\mu$ increase of $5\%$}
 \label{fig:example10}
\end{figure}

In order to be more aggressive in bond coupons, one could increase the parameters of Table~\ref{tab:example2}, namely, the loss frequency (severity) parameter $\lambda$ (resp. $\mu$). As an example, Figures~\ref{fig:example9}, \ref{fig:example10} show par yield curves for $25\%$ (resp. $5\%$) increase of $\lambda$ (resp. $\mu$) value. The resulting cyber bond par yields are then $6.52\%$ and $10.92\%$, respectively. One could tune both parameters increase according to the needs of the bond issuer. It is also easy to see that the unreasonable increase of the loss severity parameter $\mu$ could badly influence a bond coupon rate (namely, in the current cyber bond setting, $5\%$ increase of $\mu$ almost doubles the coupon rate obtained through $25\%$ increase of $\lambda$).

\par
\begin{figure}[htb]
 \centering
 \includegraphics[width=0.55\textwidth, height=0.55\textwidth]{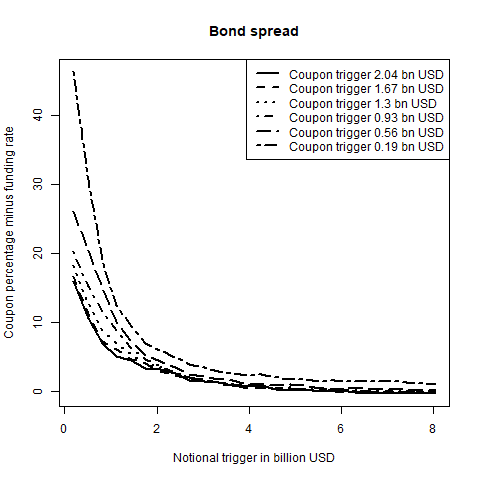}
 \caption{Cyber bond spread}
 \label{fig:example11}
\end{figure}
\begin{figure}[htb]
 \centering
 \includegraphics[width=0.55\textwidth, height=0.55\textwidth]{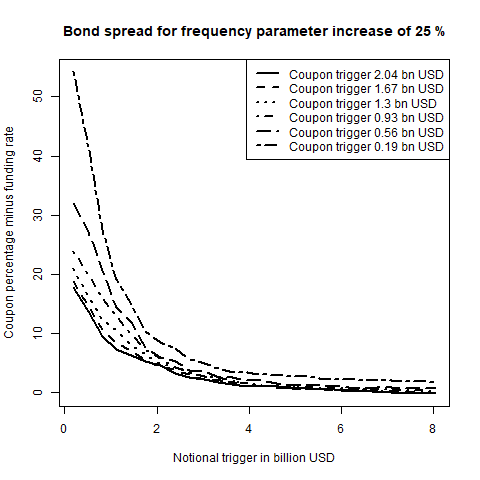}
 \caption{Cyber bond spread for loss frequency parameter $\lambda$ increase of $25\%$}
 \label{fig:example12}
\end{figure}
\begin{figure}[htb]
 \centering
 \includegraphics[width=0.55\textwidth, height=0.55\textwidth]{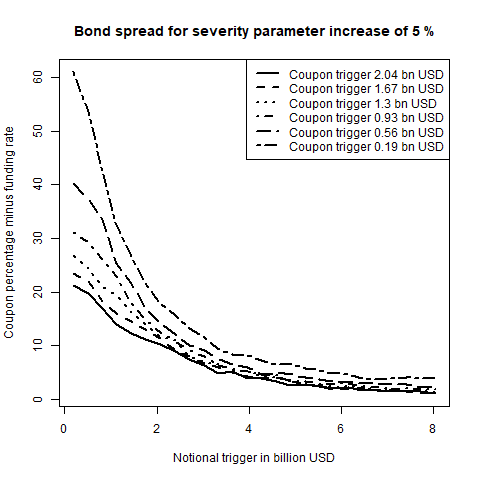}
 \caption{Cyber bond spread for loss severity parameter $\mu$ increase of $5\%$}
 \label{fig:example13}
\end{figure}

Notice that increasing the cyber loss frequency or severity parameter will influence other cyber bond characteristics as well. For example, Figures~\ref{fig:example11}, \ref{fig:example12}, and \ref{fig:example13} show cyber bond spreads (namely, bond coupon percentage minus funding rate) for initial $\lambda$ and $\mu$ as well as their increased values for $25\%$ and $5\%$, respectively.

\par
\begin{figure}[htb]
 \centering
 \includegraphics[width=0.55\textwidth, height=0.55\textwidth]{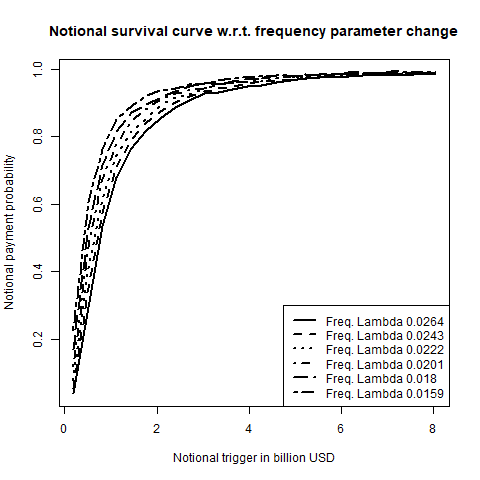}
 \caption{Cyber bond notional survival curve for $3$ years for $\lambda$ in the interval $\lambda\pm\lambda\cdot 25\%$}
 \label{fig:example14}
\end{figure}
\begin{figure}[htb]
 \centering
 \includegraphics[width=0.55\textwidth, height=0.55\textwidth]{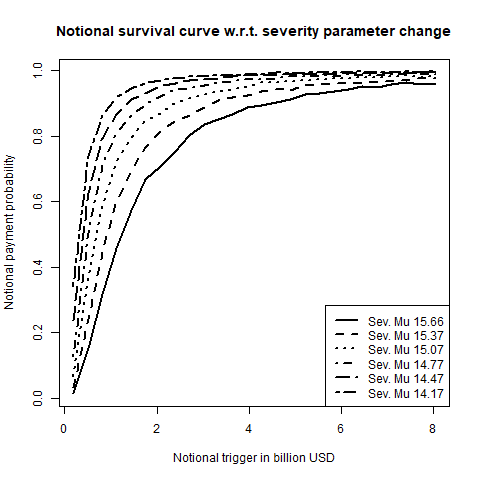}
 \caption{Cyber bond notional survival curve for $3$ years for $\mu$ in the interval $\mu\pm\mu\cdot 5\%$}
 \label{fig:example15}
\end{figure}

Additionally, Figures~\ref{fig:example14}, \ref{fig:example15} show the bond notional survival curves for loss frequency ($\lambda$) and loss severity ($\mu$) in the intervals $\lambda\pm\lambda\cdot 25\%$ and $\mu\pm\mu\cdot 5\%$, respectively. It can be easily seen from Figure~\ref{fig:example15} that $5\%$ loss severity parameter increase visibly reduces notional payment probability with the set of current possible notional triggers.

\par
\begin{figure}[htb]
 \centering
 \includegraphics[width=0.55\textwidth, height=0.55\textwidth]{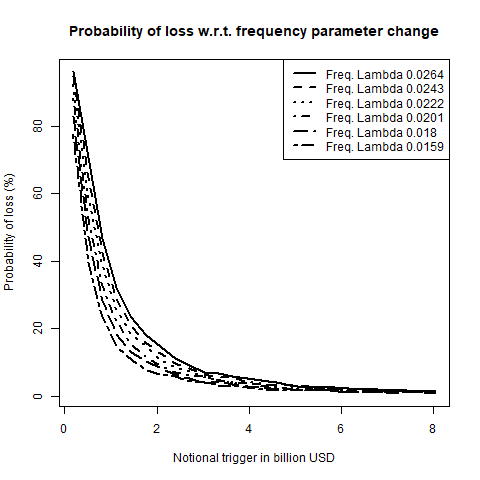}
 \caption{Cyber bond probability of loss for $3$ years for $\lambda$ in the interval $\lambda\pm\lambda\cdot 25\%$}
 \label{fig:example16}
\end{figure}
\begin{figure}[htb]
 \centering
 \includegraphics[width=0.55\textwidth, height=0.55\textwidth]{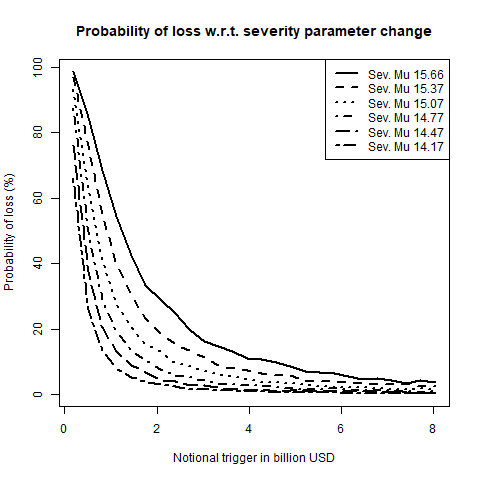}
 \caption{Cyber bond probability of loss for $3$ years for $\mu$ in the interval $\mu\pm\mu\cdot 5\%$}
 \label{fig:example17}
\end{figure}

Finally, Figures~\ref{fig:example16}, \ref{fig:example17} show the probability of (notional) loss $PL$ for different notional triggers for loss frequency ($\lambda$) and loss severity ($\mu$) in the intervals $\lambda\pm\lambda\cdot 25\%$ and $\mu\pm\mu\cdot 5\%$, respectively. These two figures just mirror the results of Figures~\ref{fig:example14}, \ref{fig:example15}.

%%% ----------------------------------------------------------------------

\subsection{Simple result analysis\label{sec:example7}}

%%% ----------------------------------------------------------------------

\par
The numerical results of Section~\ref{sec:example3} show that the choice of notional and in some cases, coupon trigger could significantly influence our proposed cyber bond parameters. For example, as follows from Figure~\ref{fig:example4}, under the small notional and coupon triggers, the cyber bond yield could reach double-digits. Such a high bond yield, however, comes from a rather low bond price as per Figure~\ref{fig:example6}. It is additionally influenced by the bond risk premium, which should be tuned to the needs of a specific investor. Under the ``reasonable" notional and coupon triggers, as follows from Figure~\ref{fig:example5}, the benefit of investing into our proposed cyber bond over relying on the funding rate is about $3\%$. Since we propose a general cyber bond framework only, it is up to the actual risk-taker to decide which notional and coupon trigger should be deemed reasonable. Moreover, other trigger types (instead of just summing up losses up to a certain date) could be explored in case it is necessary for the bond issuer as mentioned at the end of Section~\ref{sec:example1}.

\par
The two techniques to calculate a cyber bond coupon rate considered in Section~\ref{sec:example4} could provide an entry point for the actual calculations done by the bond issuer. The probability of the loss approach provides a higher coupon rate in comparison with the par yield approach ($13.63\%$ versus $5.09\%$, respectively). The latter rate, however, could be easily increased through changing the cyber loss frequency and loss severity parameters (to $6.52\%$ and $10.92\%$, respectively). Changing those parameters though could lead to a significant worsening of other cyber bond characteristics, e.g., increasing the probability of the notional loss for the notional trigger.

%%% ----------------------------------------------------------------------

\section{Conclusion}

%%% ----------------------------------------------------------------------

\par
This paper presented a general setting of cyber bonds and considered a specific and extensive cyber bond example, including a bond price, yield, risk premium, \emph{etc.} The setting appears to be convenient to use and could be easily tuned to the needs of the bond issuer. Moreover, we showed that the publicly available cyber loss event databases provide enough information to estimate the cyber loss severity and frequency distributions. Finally, two important points arose from our investigation. First, the proposed cyber bond characteristics are heavily influenced by coupon and notional triggers (determining the payment of bond coupons and notional, respectively). Second, the loss frequency distribution of specific cyber event groups can be different from that of the whole cyber event database. Both points should be necessarily cared about by the actual bond issuer.

\par
For several years, several financial institutions have been using operational risk securities that specifically cover cyber risk. The materiality of these securities is growing constantly, and the separation of their scope is a matter of time. The proposed new type of financial instruments based on cyber risk allows satisfying the investors' demand for the range of available products and extend the range of potential returns and risks. On the other hand, issuers will be able to decrease coupon payments as compared to operational risk securities and mitigate the damage from cyberattacks.

\par
The proposed financial instrument factors in the accidental occurrence of cyber incidents and unexpected losses after such events. It is considered that losses cannot be defined promptly. Thus, the modeled loss trigger is used to catch the moment of a potential cyberattack.

\par
The proposed cyber risk securities are the instruments of current interest since it is complicated to predict the underlying risk. Furthermore, we currently have no methods to assess the potential damage from a cyberattack as far as the real damage immediately after the incident.

\par
Further research is aimed at finding a method that will allow assessing losses from a cyber incident immediately. Thus, we will be able to use an indemnity trigger in the corresponding financial instrument.

%%% End document
\end{document}